\documentclass[12pt]{iopart}

\usepackage{txfonts,color}
\usepackage{graphicx}
\usepackage[sort,compress]{cite}
\usepackage{iopams}
\usepackage{setstack}
\usepackage{stmaryrd}
\usepackage{multirow,color}

\begin{document}

\title{Limit-order book resiliency after effective market orders: Spread, depth and intensity}

\author{Hai-Chuan Xu$^{1,2}$, Wei Chen$^3$, Xiong Xiong$^4$, Wei Zhang$^4$, Wei-Xing Zhou$^{1,2,5}$ and H. Eugene Stanley$^{6}$}

\address{$^1$ Department of Finance, East China University of Science and Technology, Shanghai 200237, China}
\address{$^2$ Research Center for Econophysics, East China University of Science and Technology, Shanghai 200237, China}
\address{$^3$ Shenzhen Stock Exchange, 5045 Shennan East Road, Shenzhen 518010, China}
\address{$^4$ College of Management and Economics, Tianjin University, Tianjin 300072, China}
\address{$^5$ Department of Mathematics, East China University of Science and Technology, Shanghai 200237, China}
\address{$^6$ Center for Polymer Studies and Department of Physics, Boston University, Boston, Massachusetts 02215, USA}

\ead{wxzhou@ecust.edu.cn (W.-X. Zhou)}

\begin{abstract}
  In order-driven markets, limit-order book (LOB) resiliency is an important microscopic indicator of market quality when the order book is hit by a liquidity shock and plays an essential role in the design of optimal submission strategies of large orders. However, the evolutionary behavior of LOB resilience around liquidity shocks is not well understood empirically. Using order flow data sets of Chinese stocks, we quantify and compare the LOB dynamics characterized by the bid-ask spread, the LOB depth and the order intensity surrounding effective market orders with different aggressiveness. We find that traders are more likely to submit effective market orders when the spreads are relatively low, the same-side depth is high, and the opposite-side depth is low. Such phenomenon is especially significant when the initial spread is 1 tick. Although the resiliency patterns show obvious diversity after different types of market orders, the spread and depth can return to the sample average within 20 best limit updates. The price resiliency behavior is dominant after aggressive market orders, while the price continuation behavior is dominant after less-aggressive market orders. Moreover, the effective market orders produce asymmetrical stimulus to limit orders when the initial spreads equal to 1 tick. Under this case, effective buy market orders attract more buy limit orders and effective sell market orders attract more sell limit orders. The resiliency behavior of spread and depth is linked to limit order intensity.
\end{abstract}

\submitto{J. Stat. Mech.}

\maketitle

{\color{blue}{\tableofcontents}}

\section{Introduction}
\label{s1:intro}

Resiliency is an important measure of market liquidity. A market where prices recover quickly after a liquidity shock is defined as a resilient market \cite{Kyle-1985-Em}. Now with the popularity of electronic order-driven market, the definition of resiliency is extended. A limit order book is called resilient when it reverts to its normal shape promptly after trades \cite{Large-2007-JFinM}.

Studies on LOB resiliency have been carried out on different time horizons. Many researchers focus on minutely and daily time scales. For the Swedish stock index futures, the shocks to depth are restored in less than 60 minutes\cite{Coppejans-Domowitz-Madhavan-2004}. For the NYSE and the NASDAQ stocks, the resiliency dynamics of volatility, volume and bid-ask spread are examined after experiencing a large liquidity shock\cite{Zawadowski-Andor-Kertesz-2006-QF}. Further, a power-law decay is observed for the resiliency of the bid-ask spread and the volatility\cite{Ponzi-Lillo-Mantegna-2009-PRE,Jiang-Chen-Zheng-2013-PA,Xu-Zhang-Liu-2014-PA}. In addition, the impact of institutional trading on stock resiliency during the financial crisis of 2007-2009 is also studied from a long horizon perspective \cite{Anand-Irvine-Puckett-Venkataraman-2013-JFE}.

Other researchers analyze market resiliency at order-event scales. Generally, There are two methods to conduct resiliency analysis on this shortest scale, i.e. Hawkes processes and global average measures. The first method views price changes, order submissions and cancelations as a single-variable or multi-variable Hawkes point process. The order-event intensities with a ten-variable point process model are estimated, showing that the order book does not replenish reliably after a large trade in over 60 percent of cases \cite{Large-2007-JFinM}. The trades-through model constructed by a bivariate Hawkes process suggest that the cross-exciting effect of buy and sell trades-through is weak \cite{Tock-Pomponio-2012-EEJ}. Hawkes processes can also model the resiliency of high frequency financial price jump events\cite{Bacry-Dayri-Muzy-2012-EPJB}. More interestingly, the separation that how much of price reflexivity is due to endogenous feedback processes can even be quantified by the Hawkes model \cite{Filimonov-Sornette-2012-PRE,Filimonov-Bicchetti-Maystre-Sornette-2014-JIMF,Filimonov-Wheatley-Sornette-2015-CNSNS}.

Although Hawkes process methods are more quantified, they can only characterize one dimension of liquidity, named intensity of order events. In contrast, global average measures can include different dimensions of liquidity. The evolution of depth and spreads as well as the prices and durations at the best prices around aggressive orders is investigated by the average measures\cite{Degryse-deJong-vanRavenswaaij-Wuyts-2005-RF}. They show that the depth and spread return to their initial levels within 20 best limit updates after the shock. Around large intraday price changes in the Shenzhen Stock Exchange, the volatility, the volume of orders, the bid-ask spread, and the volume imbalance decay slowly as a power law\cite{Mu-Zhou-Chen-Kertesz-2010-NJP}. Differently, There shows fast resiliency to ``normal levels'' after large transactions in \cite{Gomber-Schweickert-Theissen-2015-EFM}. In addition, all dimensions of liquidity are found to revert to their steady-state values within 15 orders after a very aggressive market order shock, based on a VAR model\cite{Wuyts-2012-EJF}.

We note that these empirical studies focus mainly on aggressive orders, usually defined as the set of market orders that move the best price, like ``large transactions'', ``trades-through'' and ``extreme order events'', etc. This paper contributes to this literature by performing multi-dimension analysis (bid-ask spread, depth and intensity) of limit order book resiliency around effective market orders. Our research differs in three ways from previous papers. First, we include all effective market orders, not only aggressive orders, considering that less aggressive orders (Type 3 and Type 9 orders below) account for the largest proportion of effective market orders. Second, seeing that Foucault et al build an equilibrium strategies model of an order-driven market and show that the spread is negatively related to the proportion of patient traders and their order arrival rate \cite{Foucault-Kadan-Kandel-2005-RFS}, which indicates predictable relation between trading intensity and spread, we try to examine empirically the relationship between spread/depth and order intensity. Third, due to various intra-day seasonality, we remove the intra-day seasonality using the Fourier Flexible Form (FFF) regression before the global average measure.

\section{Materials and methods}

\subsection{Dimensions of limit-order book resiliency}


We describe the limit order book first. The order book right before the $t$-th event can be described as follows
\begin{equation}
 \begin{array}{cccccccccccccccccc}
  \cdots, & b_2, & b_1, & a_1, & a_2, \cdots \\%
  \cdots, & B_2, & B_1, & A_1, & A_2, \cdots   %
 \end{array}
 \label{Eq:BidLOB:AB}
\end{equation}
where $b_i$ and $a_i$ are respectively the $i$-th bid and ask prices
and $B_i$ and $A_i$ are the associated volumes at the corresponding
quotes.

Kyle defines market liquidity along three dimensions: (i) tightness, ``the cost of turning around a position over a short period of time'', measured by bid-ask spread $\tilde{s}(t) = a_1(t)-b_1(t)$, (ii) depth, ``the size of an order flow innovation required to change prices a given amount'', measured by pending volume at the best quotes ($B_1$ and $A_1$) if the given amount is 1 tick, and (iii) resiliency, ``the speed with which prices recover from a random, uninformative shock'' \cite{Kyle-1985-Em}. For order-driven markets, the definition of resiliency is extended as the speed with which the LOB reverts to its normal shape. Hence, LOB resiliency after shocks can be characterized by the evolution of bid-ask spread, depth and intensity, which is defined by the expected number of events in a unit time interval.

\subsection{Data sets}

Our data sets were extracted from the order flow data of 31 A-share stocks and 12 B-share stocks traded on the Shenzhen Stock Exchange in 2003. The key distinction is that A-shares are denominated in Renminbi and B-shares in Hong Kong Dollar. The A-shares market was open only to domestic investors in 2003. The market consists of three time periods on each trading day, namely, the opening call auction, the cooling period, and the continuous double auction period. Here we only consider the order flow occurring in the continuous double auction period (9:30 AM to 11:30 AM and 1:00 PM to 3:00 PM, 240 mins for each day). Because the results for different stocks are very similar, we only present the results of one A-share stock (000858) and one B-share stock (200002).

\subsection{Order types}

We present the classification of orders. Assume that, right before the arrival of an effective market order, the sequences of prices and volumes on the bid side of the LOB are $\{b_i: i=1,2,\cdots\}$ and $\{B_i: i=1,2,\cdots\}$ and those on the ask side are $\{a_i: i=1,2,\cdots\}$ and $\{A_i: i=1,2,\cdots\}$, respectively. Without loss of generality, assume that $b_m<\cdots<b_2<b_1 <a_1<a_2<\cdots<a_n$, where $b_m$ and $a_n$ are respectively the minimal bid price and maximal ask price. These four sequences determine the current status of the LOB right before the arrival of an effective market order.

Consider an effective market order of price $\pi$ and size $\omega$. This order can be decomposed into two parts, the executed part and the remaining part that is not executed, such that
\begin{equation}
 \omega = \omega_e + \omega_r~,
 \label{Eq:LOB:Resiliency:omega:er}
\end{equation}
where $\omega_e$ is the size of the executed part and $\omega_r$ is the size of the remaining part. When $\omega_r=0$, all shares of the order are filled. This type of orders is termed filled effective market orders, or {\em{filled orders}} for short. When $\omega_r\neq0$, only part of the order is filled and we can call this type of orders as partially filled effective market orders, or {\em{partially filled orders}} for short. Zhou has shown that partially filled orders have remarkably larger market impact than filled orders. Hence, partially filled orders are more aggressive than filled orders \cite{Zhou-2012-QF}.

Many empirical studies measure order aggressiveness by the position of the order price relative to that of the latest best quotes \cite{Biais-Hillion-Spatt-1995-JF,Griffiths-Smith-Turnbull-White-2000-JFE,Ranaldo-2004-JFinM,Tseng-Chen-2016-PBJF}. However, more precisely, the aggressiveness of an effective market order can also be partly captured by its penetrability. The penetrability $p$ of an effective market order can be defined as the number of levels on the opposite LOB side it consumes:
\begin{equation}
 p=\min_j\{j: \omega_e \leqslant A_1+\cdots+A_j\}
 \label{Eq:LOB:Resiliency:Penetrability:A}
\end{equation}
for an effective buy market order and
\begin{equation}
 p=\min_j\{j: \omega_e \leqslant B_1+\cdots+B_j\}
 \label{Eq:LOB:Resiliency:Penetrability:B}
\end{equation}
for an effective sell market order. Therefore, effective market orders are orders with $p\geqslant1$ and effective limit orders are characterized by $p=0$. We find that, for an effective market order of penetrability $p$, $A_1+\cdots+A_{p-1}<\omega \leqslant A_1+\cdots+A_p$ and $\pi\geqslant a_p$ for filled buy orders, $\omega>A_1+\cdots+A_{p}$ and $\pi= a_p$ for partially filled buy orders, $B_1+\cdots+B_{p-1}<\omega \leqslant B_1+\cdots+B_p$ and $\pi\leqslant b_p$ for filled sell orders, and $\omega > B_1+\cdots+B_p$ and $\pi= b_p$ for partially filled sell orders.

We utilize the classification scheme consistent with that of \cite{Degryse-deJong-vanRavenswaaij-Wuyts-2005-RF}. Degryse et al. classify incoming orders into 12 types based on order price and penetrability \cite{Degryse-deJong-vanRavenswaaij-Wuyts-2005-RF}. Specifically, the effective buy (and sell) market orders are classified into three types (types 1, 2 and 3 for buys and types 7, 8 and 9 for sells). Orders of Type-1 in \cite{Degryse-deJong-vanRavenswaaij-Wuyts-2005-RF} corresponds to bid orders with the order size greater than the best ask size and the order price higher than the best ask price. In our terminology, a Type-1 order is an effective buy market order whose penetrability $p$ is greater than 1. This means that these orders walk up the limit order book and result in multiple trades. An order of Type-2 in \cite{Degryse-deJong-vanRavenswaaij-Wuyts-2005-RF} is an order with the size greater than the best ask size, but it does not walk up the limit order book above the best ask. In other words, a Type-2 order is a partially filled order with $p=1$. Orders of Type-3 in \cite{Degryse-deJong-vanRavenswaaij-Wuyts-2005-RF} are bid market orders whose order size is lower than the best ask size. Speaking differently, a Type-3 order is a filled order with $p=1$. Similarly, Type-7 orders are effective sell orders with $p>1$, Type-8 orders are partially filled sell orders with $p=1$, and Type-9 orders are filled sell with $p=1$. Obviously, orders of types 1, 2 and 3 form the whole set of effective buy market orders and those of types 7, 8 and 9 constitute the whole set of effective sell market orders. \cite{Degryse-deJong-vanRavenswaaij-Wuyts-2005-RF} focus on the orders of types 1, 2, 7 and 8.

\begin{figure}[!tb]
 \centering
\includegraphics[width=0.6\textwidth]{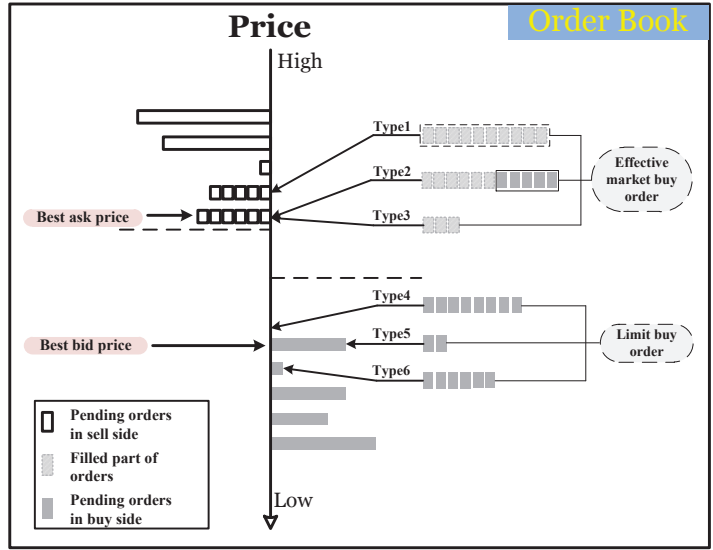}
\caption{Classification of buy orders. The order prices are pointed by arrows. The number of small rectangles indicates order size.}
\label{Fig:Orderbook}
\end{figure}

The remaining order types are not executed immediately. The prices of Type-4 orders are lower than the best ask, but higher than the best bid price, that is, Type-4 orders are buy limit orders placed in the spread. The prices of Type-5 orders are exactly at the best bid. The prices of Type-6 orders are lower than the best bid. Symmetrically, Type-10 orders are sell limit orders placed in the spread; Type-11 orders are sell limit orders placed exactly at the best ask; The remaining orders are Type-12 orders. Fig.~\ref{Fig:Orderbook} illustrates all the 6 types of orders on the buy side.

Table~\ref{Tb:TypeData} shows the numbers of orders of the 12 types for stocks 000858 and 200002. We can find that Type-3 orders and Type-9 orders are most popular in the effective market orders. In addition, for all types of orders, the vast majority are placed when the spread is 1 tick.

\begin{table}[!bp]
 \centering
 \caption{\label{Tb:TypeData} The numbers of 12 different types of orders for stock 000858 and 200002 with different values of bid-ask spreads.}
 \medskip
 \begin{tabular}{crrrrrrrrr}
  \hline\hline
  \multirow{3}*[3mm]{Type} &  \multicolumn{4}{c}{000858} && \multicolumn{4}{c}{200002} \\
     \cline{2-5} \cline{7-10}
  & $s=0.01$ & $s=0.02$ & $s=0.03$ & $s\geqslant0.04$ && $s=0.01$ & $s=0.02$ & $s=0.03$ & $s\geqslant0.04$ \\\hline
 Type 1          & 9346 & 1981 & 635 & 421 && 1688 & 467 & 162 & 203  \\
 Type 2          & 12568 & 1801 & 442 & 232 && 3673 & 598 & 152 & 116  \\
 Type 3          & 97038 & 14916 & 3735 & 2473 && 16026 & 2795 & 782 & 682  \\
 Type 4          & 0 & 8302 & 3370 & 2957 && 0 & 1934 & 993 & 1265  \\
 Type 5          & 69485 & 14331 & 4536 & 3189 && 12844 & 3858 & 1384 & 1458  \\
 Type 6          & 135210 & 32123 & 10145 & 7281 && 23297 & 7282 & 2740 & 3441  \\
 Type 7          & 10593 & 2311 & 672 & 447 && 1764 & 507 & 192 & 189  \\
 Type 8          & 12063 & 1935 & 451 & 252 && 3751 & 716 & 211 & 140  \\
 Type 9          & 90510 & 13412 & 3346 & 1815 && 15378 & 2718 & 771 & 657  \\
 Type 10          & 0 & 7520 & 3065 & 2487 && 0 & 1600 & 807 & 1066  \\
 Type 11          & 61318 & 12451 & 3474 & 2343 && 11722 & 3117 & 982 & 1215  \\
 Type 12          & 192156 & 44715 & 13863 & 9711 && 30682 & 8913 & 3352 & 4088  \\
  \hline\hline
 \end{tabular}
\end{table}


\subsection{Methods}

To quantify the limit order book resiliency, we average the spread, depth and intensity around effective market orders of the same type and compare the results among different types. Before doing the averaging measure, we should remove the intra-day seasonality of these 3 resiliency proxies. We uses the Fourier Flexible Form (FFF) regression framework introduced by Gallant \cite{Gallant-1981-JEm} to characterize periodic patterns, which has been applied for modelling intra-day periodic returns, volatilities and quantities of financial securities \cite{Andersen-Bollerslev-1997-JEF,Andersen-Bollerslev-Das-2001-JF,Hardle-Hautsch-Mihoci-2012-JEF}. We employ the FFF framework to regress the intra-day periodic component $x(\tau)\in\{s(\tau), d(\tau), \lambda(\tau)\}$ for spread $\tilde{s}$, depth at best bid/ask $\tilde{d}$ and intensity $\tilde{\lambda}$. Note that this paper only pay attention to the intensity of Type-4, Type-5, Type-10 and Type-11 orders, considering that the limit order book resiliency after an effective market order is mainly achieved by these 4 types orders. The intra-day periodic component can be expressed as follows,
\begin{equation}
 x(\tau)=\sum^{Q}_{q=0}\alpha_q\left(\frac{\tau}{T}\right)^q + \sum^{P}_{p=1}\left[\beta_{c,p}\cos\left(2\pi p\frac{\tau}{T}\right) + \beta_{s,p}\sin\left(2\pi p\frac{\tau}{T}\right)\right],
 \label{Eq:Liquidity:Seasonality}
\end{equation}
where $\tau$ presents the $\tau$-th intra-day interval and $T=240$ is the maximum number of the 1-min intervals. The parameters $Q$ and $P$ are the orders of the Fourier expansion. Andersen et al. suggest that the Fourier terms in Eq.~(\ref{Eq:Liquidity:Seasonality}) beyond $Q=2$ and $P=6$ produce insignificant estimates for any additional $\alpha_q$, $\beta_{c,p}$, and $\beta_{s,p}$ coefficients \cite{Andersen-Bollerslev-Cai-2000}. We thus choose $Q=2$ and $P=6$. The regression will produce the unique set of the periodic intra-day estimates $x(1),x(2),...,x(240)$.

\begin{figure}[!tb]
  \centering
  \includegraphics[width=0.99\linewidth]{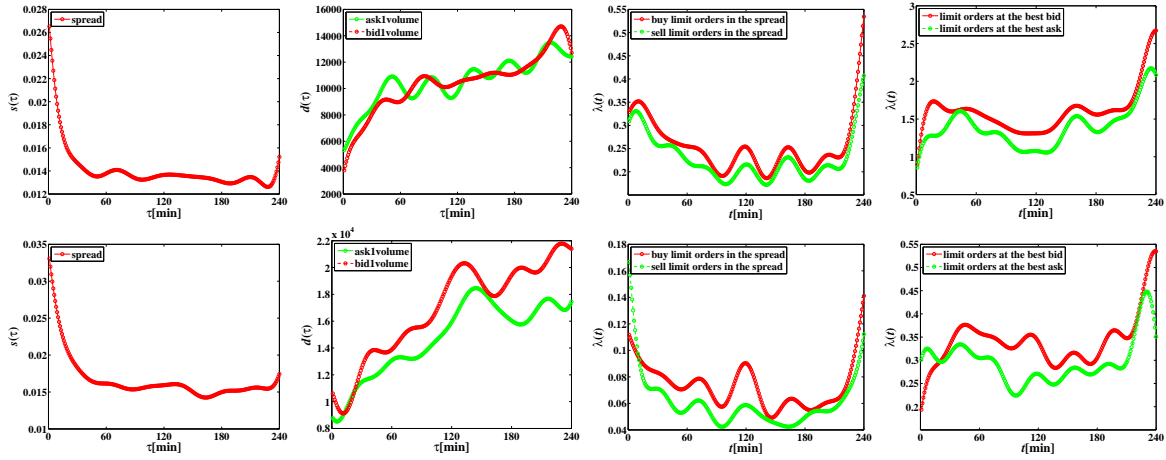}
  \caption{(Color online) Estimated 1-min intra-day seasonality factors for 000858 (top panel) and 200002 (bottom panel) traded from January to December, 2003. The columns from left to right correspond to the intra-day seasonality patterns of the bid-ask spread $s(\tau)$, the depth $d(\tau)$, the intensity $\lambda(\tau)$ of Type-4/10 orders, and the intensity $\lambda(\tau)$ of Type-5/11 orders.}
  \label{Fig:LOBResilency:Seasonality}
\end{figure}

Fig.~\ref{Fig:LOBResilency:Seasonality} displays the estimated 1-min intra-day seasonality for bid-ask spread, depth, intensity of Type-4/10 orders and intensity of Type-5/11 orders. We can observe that the intra-day seasonality of bid-ask spread appears a reversed $J$-shaped pattern, which is consistent with many other limit-order markets. On the contrary, due to the accumulation of limit orders, the volumes at the best quotes show an increasing trend. For the intensities of limit orders in the spread, their seasonality generally show a $U$-shaped pattern, indicating that higher order-in-spread intensity emerge near the opening time and the closing time. Note that the time at 120th min is also a opening/closing time. We also observe that the intensity of buy orders and sell orders fluctuates synchronously.

In order to account for the intra-day seasonality effects, we adjust the spread and depth correspondingly as follows,
\begin{equation}
 S(t)=\frac{100\langle{\tilde{s}(t)/s(\tau_t)}\rangle}{\langle{\tilde{s}(0)/s(\tau_0)}\rangle},   t=\rm{-20,-19,\ldots,19,20}
 \label{Eq:Liquidity:Spread}
\end{equation}
\begin{equation}
 D(t)=\frac{100\langle{\tilde{d}(t)/d(\tau_t)}\rangle}{\langle{\tilde{d}(0)/d(\tau_0)}\rangle},   t=\rm{-20,-19,\ldots,19,20}
 \label{Eq:Liquidity:Depth}
\end{equation}
where $t$ means the best limit updates around the effective market order. The best limit updates are defined as the updates of either the best quotes or the depth at these quotes (or a combination of both). Time $t=0$ corresponds to the spread/depth just before the effective market order. $\tau_t$ means the 1-min interval in which the $t$-th best limit update occurs. This adjustment also includes a normalized process compared with the value at time $t=0$, and we set the average value as 100 at $t=0$.

The adjustment for intensity is slightly different:
\begin{equation}
 \Lambda([t])=\langle{2\tilde{\lambda}([t])/(\lambda(\tau^-_{[t]})+\lambda(\tau^+_{[t]}))}\rangle,   t=\rm{-30,-29,\ldots,-1,1,\ldots,29,30},
 \label{Eq:Liquidity:Intensity}
\end{equation}
where [$t$] presents the $t$-th 1-min interval away from the event time $t=0$. Because [$t$] may intersect with two neighboring $\tau$-intervals, we average two $\lambda$'s of the covered intervals ($\tau^-_{[t]}$ and $\tau^+_{[t]}$) as the intra-day seasonality. Note that, for intensity, neither $\tilde{\lambda}$ nor $\Lambda$ has definition on $t=0$, because intensity must be defined over an interval.

\section{LOB resiliency analysis}

\subsection{Resiliency of bid-ask spread}

\begin{figure}[!tbp]
  \centering
  \includegraphics[width=0.66\linewidth]{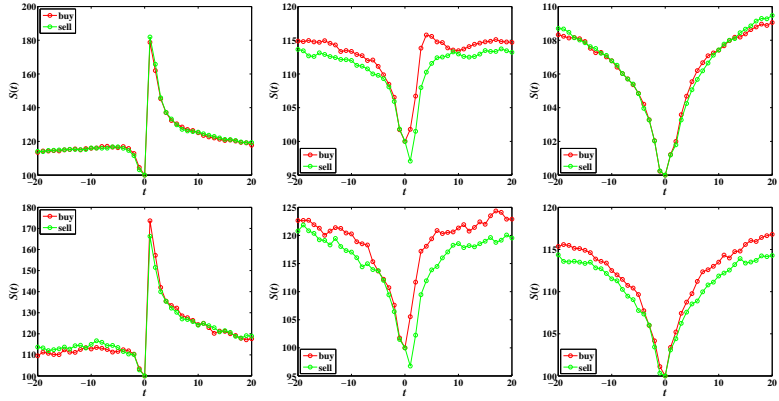}
  \caption{\label{Fig:LOBResilency:Spread} (Color online) LOB resiliency behavior of bid-ask spread around different types of effective market orders for stock 000858 (top) and 200002 (bottom). The market orders are of Type-1 (buy orders with the penetrability $p>1$)  and Type-7 (sell orders with the penetrability $p>1$) in the left panel, of Type-2 (partially filled buy orders with $p=1$) and Type-8 (partially filled sell orders with $p=1$) in the middle panel, and of Type-3 (filled buy orders with $p=1$) and Type-9 (filled sell orders with $p=1$) in the right panel. Time $t=0$ corresponds to the status of the LOB right before the arrival of an effective market order.}
\end{figure}

Fig.~\ref{Fig:LOBResilency:Spread} illustrates the average resiliency behavior of the bid-ask spread before and after the six types of effective market orders with different aggressiveness. The first feature is that the resilience behaviors for the A-share stock and the B-share stock are very similar. By definition, the relative spread is 100 right before the effective market orders. We notice that the relative spread $S(0^-)$ is approximately minimal in almost all cases, which indicates that submitting market orders are more likely when the liquidity is high. What is intriguing is the obvious difference between the resiliency behavior for different types of orders.

In the left panel of Fig.~\ref{Fig:LOBResilency:Spread}, we show the results for buy market orders with the penetrability $p>1$ (Type-1) and sell market orders with the penetrability $p>1$ (Type-7). Before the microscopic liquidity shock, the bid-ask spread increases slightly and then decreases. An effective market order of Type-1 or Type-7 consumes at least all the orders on the opposite best and the bid-ask spread soars abruptly to a peak. On average, subsequent orders are less aggressive, be they limit orders or market orders. The spread narrows gradually and relaxes to its normal level after about 20 incoming orders. The evolutionary trajectories of bid-ask spread almost overlap around buy market orders and sell market orders.

In the middle panel of Fig.~\ref{Fig:LOBResilency:Spread}, we show the results for partially filled buy market orders with the penetrability $p=1$ (Type-2) and partially filled sell market orders with the penetrability $p=1$ (Type-8). About 20 orders before the microscopic liquidity shock, the bid-ask spread starts to decrease with an acceleration when approaching to $t=0$. Right after the partially filled order at $t=0^+$, the spread narrows further. When the shock is from a buy order, the spread starts to resile and comes back to the normal level in a few updates. On the contrary, when the shock is from a sell order, the first subsequent order further narrows the spread and the resiliency begins since the second subsequent order. This phenomenon of buy-sell asymmetry is also observed for other A-share and B-share stocks.

In the right panel of Fig.~\ref{Fig:LOBResilency:Spread}, we show the results for filled buy market orders with the penetrability $p=1$ (Type-3) and filled sell market orders with the penetrability $p=1$ (Type-9). The spread decreases before the microscopic liquidity shock and increases immediately after the shock. The spread at $t=0$ is only slightly narrower than that at $t=-1$, which is due to the fact that the size of the order at $t=0^+$ is no larger than the standing volume on the opposite best. The resilience speed after orders of Type-3 and Type-9 is slower than those in the middle panel. This observation is not surprising because the first price gap between the best price and the second best price after orders of Type-2 and Type-8 is much larger than the price gap after the orders of Type-3 and Type-9 \cite{Zhou-2012-NJP,Zhou-2012-QF}.

\begin{figure}[!htb]
  \centering
  \includegraphics[width=0.99\linewidth]{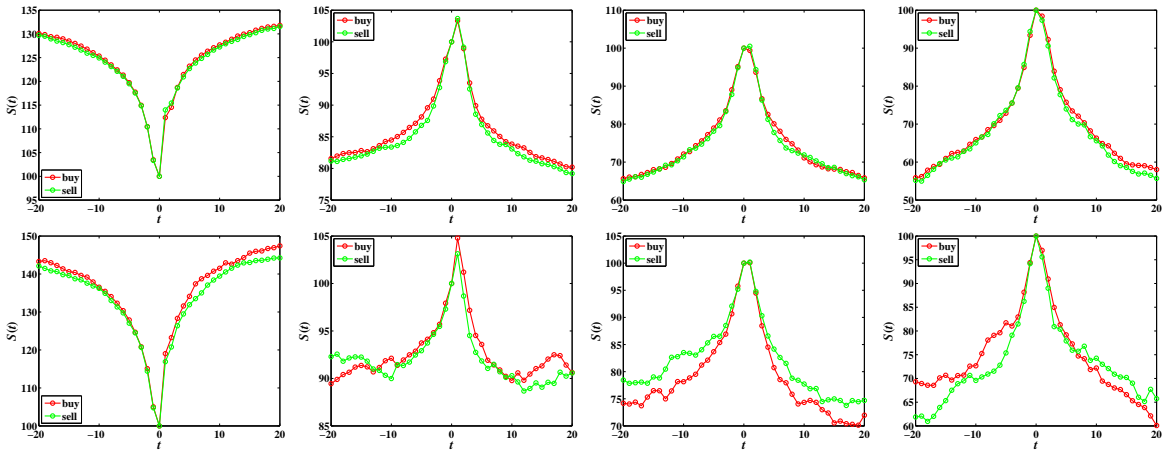}
  \caption{\label{Fig:LOBResilency:Spread:DifferentSpread} (Color online) LOB resiliency behavior of bid-ask spread around effective market orders with different initial spreads at $t=0^-$ for stock 000858 (top) and 200002 (bottom). The plots in the four columns from left to right correspond to $s(0^-)=0.01$, $s(0^-)=0.02$, $s(0^-)=0.03$ and $s(0^-)\geq0.04$. Time $t=0$ corresponds to the status of the LOB right before the arrival of an effective market order.}
\end{figure}

Fig.~\ref{Fig:LOBResilency:Spread:DifferentSpread} shows the evolution of bid-ask spread around market orders for different spreads at $t=0^-$. Because the average spread is between 0.01 CNY (1 tick) and 0.02 CNY (2 ticks), $S(t)$ decreases before the market orders at $t=0$ and increases afterwards, as shown in the left column. For larger $s(0^-)$, the bid-ask spread increases first and then decreases. For $s(0^-)=0.02$, the market orders further widen the spread and the maximum spread is at $t=1^-$. For $s(0^-)=0.03$, the market orders at $t=0$ have minor impacts on the bid-ask spread and $s(0^-)\approx s(1^-)$. For $s(0^-)\geq0.04$, the maximum spread is at $t=s(0^-)$ and the market orders usually narrow the spread, indicating that under this case, the greater falls of spread caused by Type-2 and Type-8 orders dominate the spread expansions caused by Type-1 and Type-7 orders on average, even though the absolute number of Type-2/8 orders is smaller than that of Type-1/7 orders.


\subsection{Resiliency of LOB depth}

\begin{figure}[!htb]
  \centering
  \includegraphics[width=0.99\linewidth]{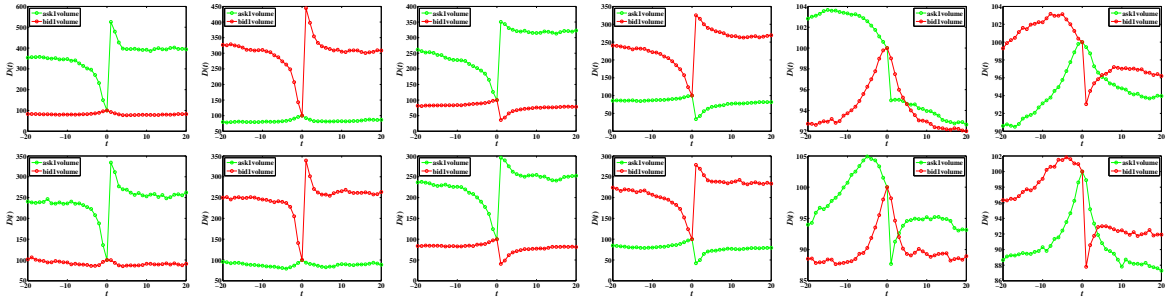}
  \caption{\label{Fig:LOBResilency:Depth:Type} (Color online) LOB resiliency behavior of depth at the best bid (red line) and the best ask (green line) around different types of effective market orders for stock 000858 (top) and 200002 (bottom). The plots in the six columns from left to right correspond to the orders of Type-1 (buy orders with the penetrability $p>1$), Type-7 (sell orders with the penetrability $p>1$), Type-2 (partially filled buy orders with $p=1$), Type-8 (partially filled sell orders with $p=1$), Type-3 (filled buy orders with $p=1$) and Type-9 (filled sell orders with $p=1$). Time $t=0$ corresponds to the status of the LOB right before the arrival of an effective market order.}
\end{figure}

Fig.~\ref{Fig:LOBResilency:Depth:Type} shows the average resiliency behavior of depth at the best quotes around the six types of effective market orders with different aggressiveness. Both the A-share stock and the B-share stock display similar resilience behaviors. We can find that the same side depth increases and the opposite side depth decreases before time $t=0$ in almost all cases, which means that effective market orders are more likely to take place when the same side depth is high and the opposite side depth is low. This is due to the consistence of market orders. Like the analysis of bid-ask spread above, what is interesting is the obvious difference between the resiliency behavior for different types of orders.

In the first and second columns of Fig.~\ref{Fig:LOBResilency:Depth:Type}, we show the results for buy market orders with the penetrability $p>1$ (Type-1) and sell market orders with the penetrability $p>1$ (Type-7), indicating the symmetry between these two types of orders. An effective market order of Type-1 or Type-7 penetrates at least one price level on the opposite order book and the neighboring level bears the new depth at the best. Empirical analysis has shown that the shape function of limit order book increases first and then decreases with respect to the distance of the price level to the best price for the Chinese stocks \cite{Gu-Chen-Zhou-2008c-PA}, which has been also observed in other markets \cite{Challet-Stinchcombe-2001-PA,Maslov-Mills-2001-PA,Weber-Rosenow-2005-QF,Bouchaud-Mezard-Potters-2002-QF,Potters-Bouchaud-2003-PA,Eisler-Kertesz-Lillo-2007-PSPIE,Gould-Porter-Williams-McDonald-Fenn-Howison-2013-QF}.
Hence, the new depth on the opposite side right after the market order of Type-1 or Type-7 is markedly higher than the normal value, while the same-side depth does not change. After the shock, the depths on the same side and the opposite side will reverse to their normal values within about 20 best limit updates.

In the third and fourth columns of Fig.~\ref{Fig:LOBResilency:Depth:Type}, we show the results for partially filled buy market orders with the penetrability $p=1$ (Type-2) and partially filled sell market orders with the penetrability $p=1$ (Type-8), indicating the symmetry between these two types of orders. These effective market orders also lead to over-resiliency of the opposite depth. This is because the depth on the second best level before the shock becomes the depth on the best price level after the shock. More interestingly, Type-2 and Type-8 orders shocks can cause over-resiliency at the same side depth. This is because the unexecuted part of the order will reside on the order book forming the new same-side best. The depths at both sides relax to their normal levels after about 20 best limit updates.

In the fifth and sixth columns of Fig.~\ref{Fig:LOBResilency:Depth:Type}, we show the results for filled buy market orders with the penetrability $p=1$ (Type-3) and filled sell market orders with the penetrability $p=1$ (Type-9). The patterns around these two types of orders are generally symmetric, except for the after-shock dynamics of opposite depth for stock 000858. Immediately after $t=0$, the opposite side depth shows a sharp decline, which reflects the liquidity consumption by an effective market order of Type-3 or Type-9. For stock 000858, after the shock of Type-3 orders, the opposite side depth continues to decline; while after the shock of Type-9 orders, the opposite side depth starts its resiliency slowly. However, this asymmetry does not appear for stock 200002, that is, the opposite side depths show a reverse pattern consistently after either a Type-3 order's shock or a Type-9 order's shock. This indicates that the asymmetry for 000858 maybe just a special case. As for the same side depths, they all gradually reverse to their normal levels like other cases.

\begin{figure}[tbp]
  \centering
  \includegraphics[width=0.99\linewidth]{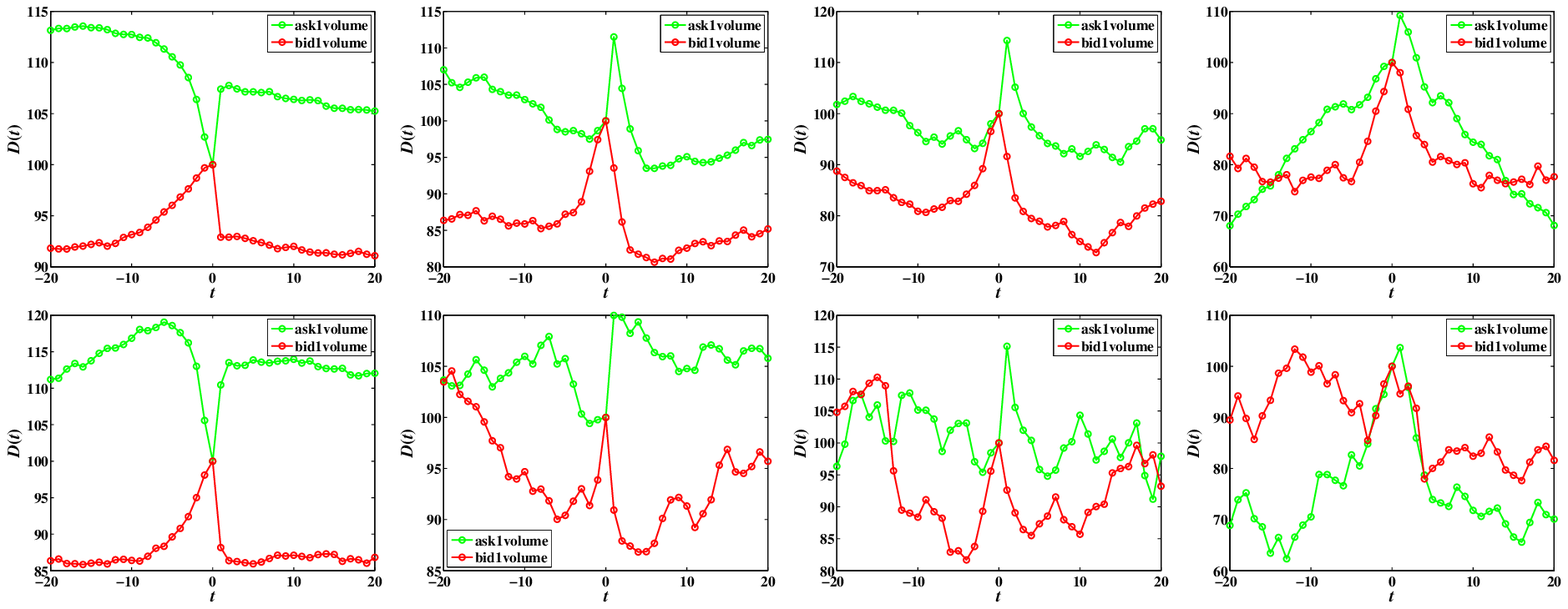}
  \includegraphics[width=0.99\linewidth]{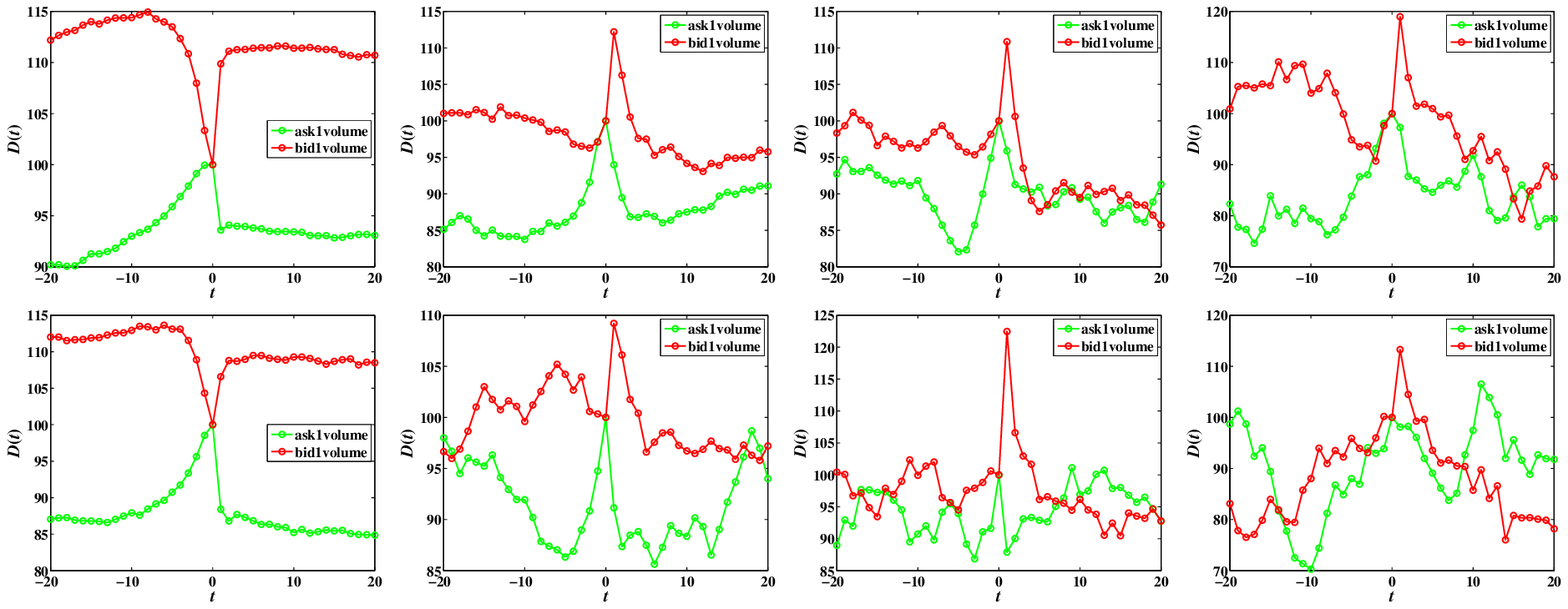}
  \caption{\label{Fig:LOBResilency:Depth:DifferentSpread} (Color online) LOB resiliency behavior of depth at the best bid (red line) and the best ask (green line) around effective market orders with different initial spread at $t=0^-$ for stock 000858 and 200002. The plots in the four columns from left to right correspond to $s(0^-)=0.01$, $s(0^-)=0.02$, $s(0^-)=0.03$ and $s(0^-)\geq0.04$. The plots in the top two rows correspond to the dynamics around effective buy market orders for stock 000858 and 200002. The plots in the bottom two rows correspond to the dynamics around effective sell market orders for stock 000858 and 200002. Time $t=0$ corresponds to the status of the LOB right before the arrival of an effective market order.}
\end{figure}

Fig.~\ref{Fig:LOBResilency:Depth:DifferentSpread} shows the evolution of depth around effective market orders for different spreads at $t=0^-$. For the opposite side depths in each plot, they show the combined effects of over-resiliency (Type-1, Type-7, Type-2 or Type-8) and sharp declining (Type-3 or Type-9). However, these combined effects are different for different spreads at $t=0^-$. For $s(0^-)=0.01$, they show full-resiliency or partial-resiliency immediately after the effective market orders and then present flat trends. For $s(0^-)>0.01$, they show over-resiliency immediately after the effective market orders and then present relaxation trends. For the same side depths in each plot, when $s(0^-)=0.01$, the market orders cause sharp decline; when $s(0^-)>0.01$, the market orders cause gradually decline. This indicates that orders of Type-2 and Type-8 are more likely submitted when $s(0^-)=0.01$, which is consistent with  Table~\ref{Tb:TypeData}. We should also notice that, the phenomenon that effective market orders take place when the same side depth is high and the opposite side depth is low, can get the best expression when the initial spread is 0.01. In other words, the first column of Fig.~\ref{Fig:LOBResilency:Depth:DifferentSpread} is significantly different from the others, which is also confirmed by Fig.~\ref{Fig:LOBResilency:Spread:DifferentSpread} for spreads and Fig.~\ref{Fig:LOBResilency:Intensity:DifferentSpread} for intensity.

\subsection{Resiliency of order intensity}

\begin{figure}[!htb]
  \centering
  \includegraphics[width=0.24\linewidth]{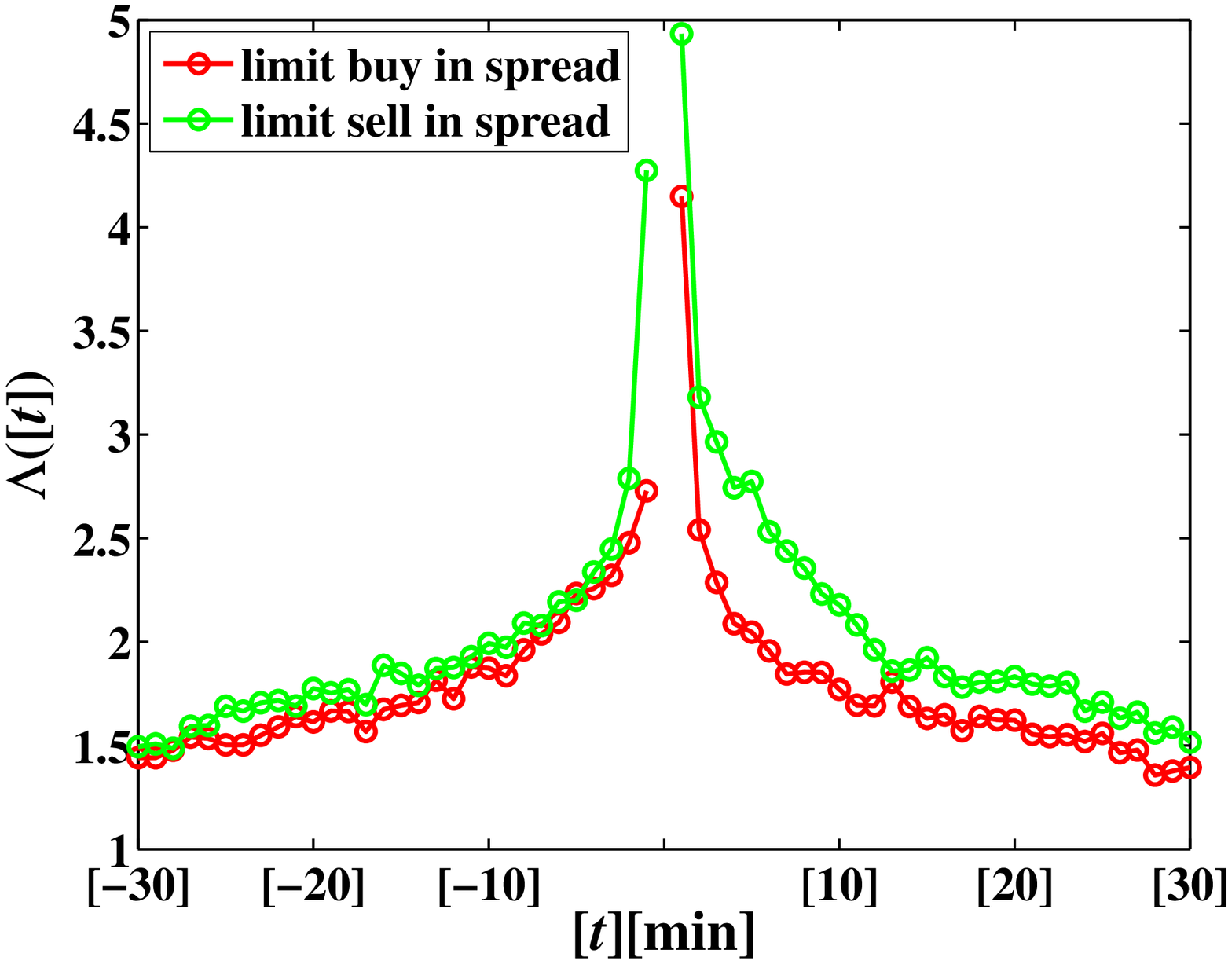}
  \includegraphics[width=0.24\linewidth]{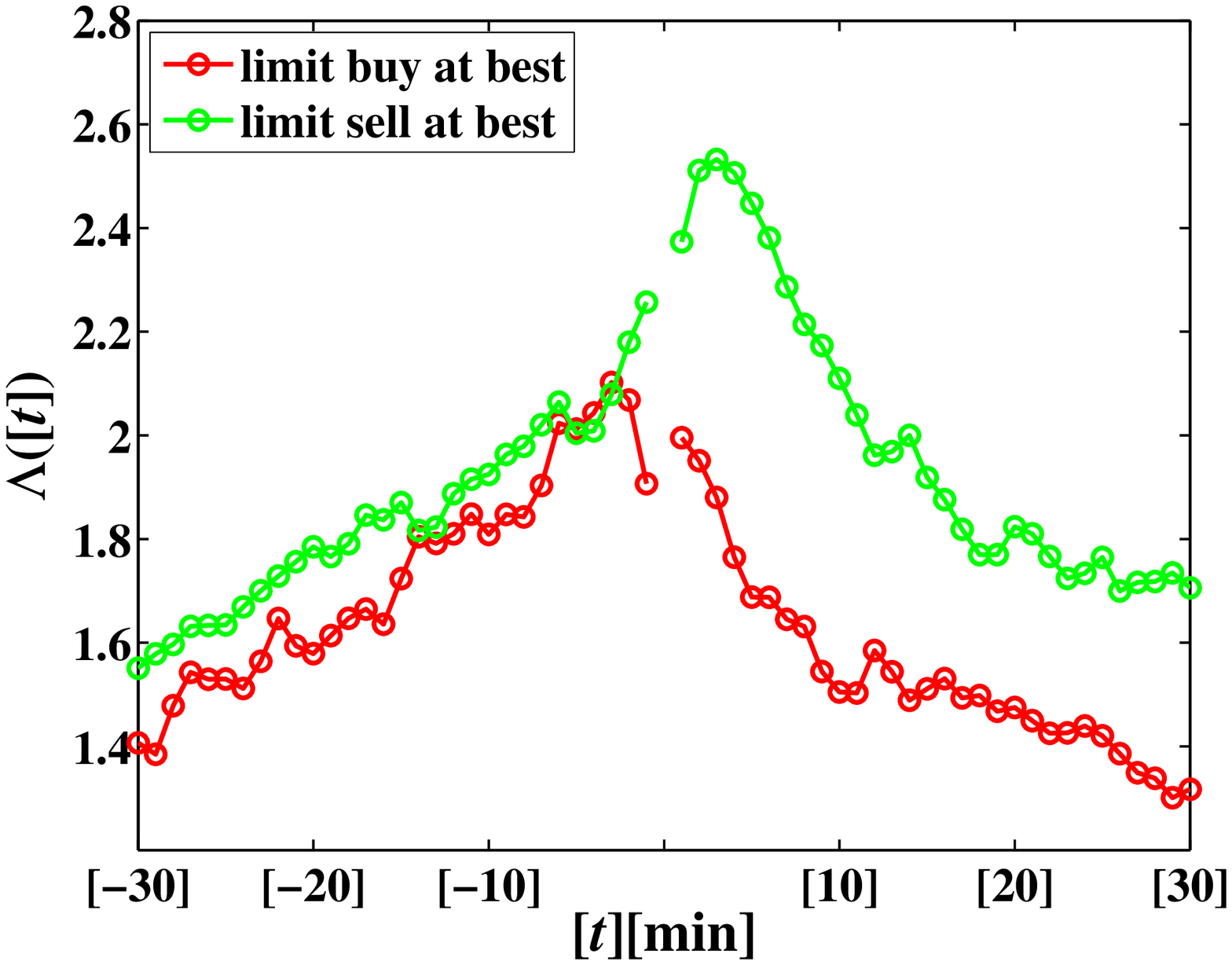}
  \includegraphics[width=0.24\linewidth]{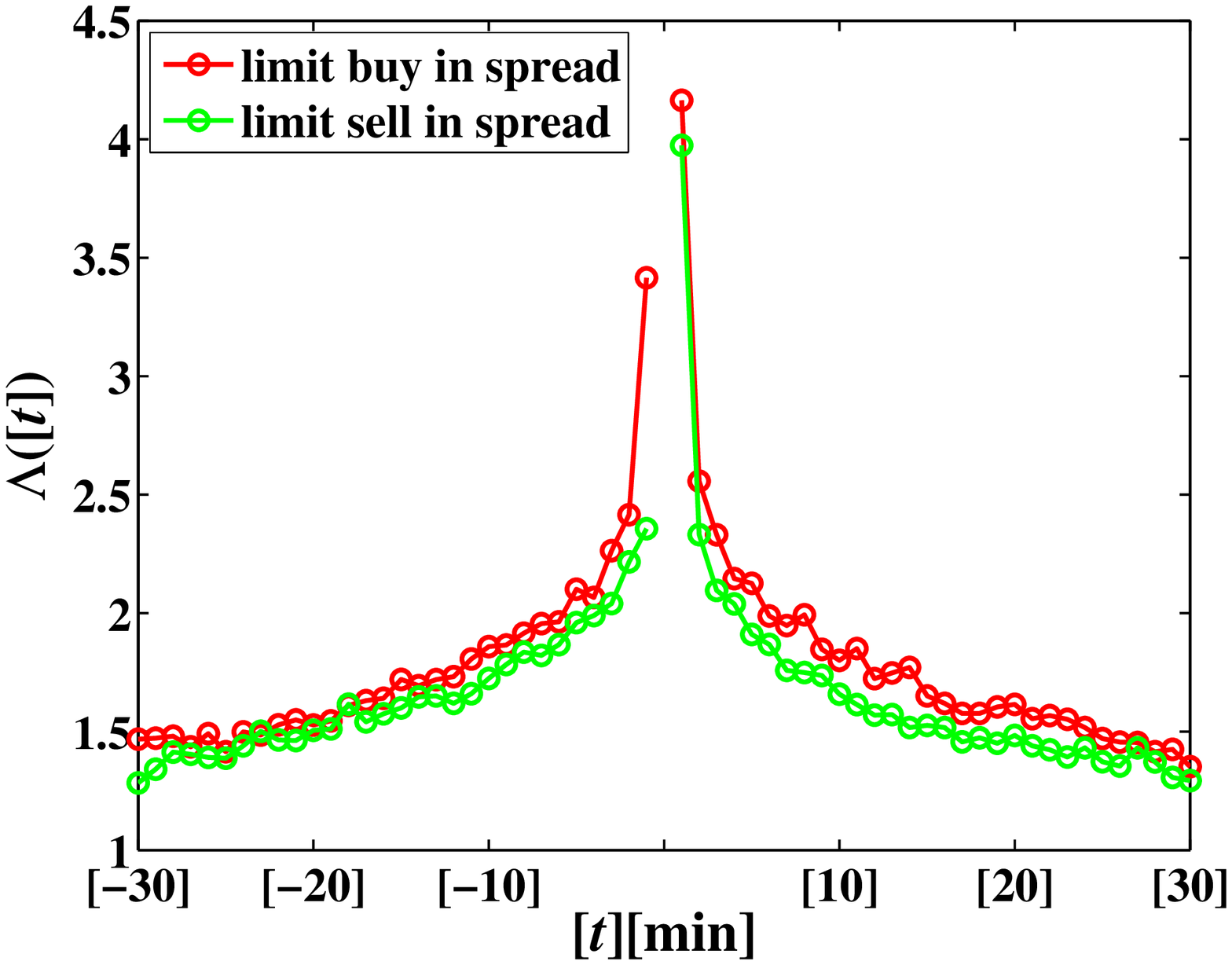}
  \includegraphics[width=0.24\linewidth]{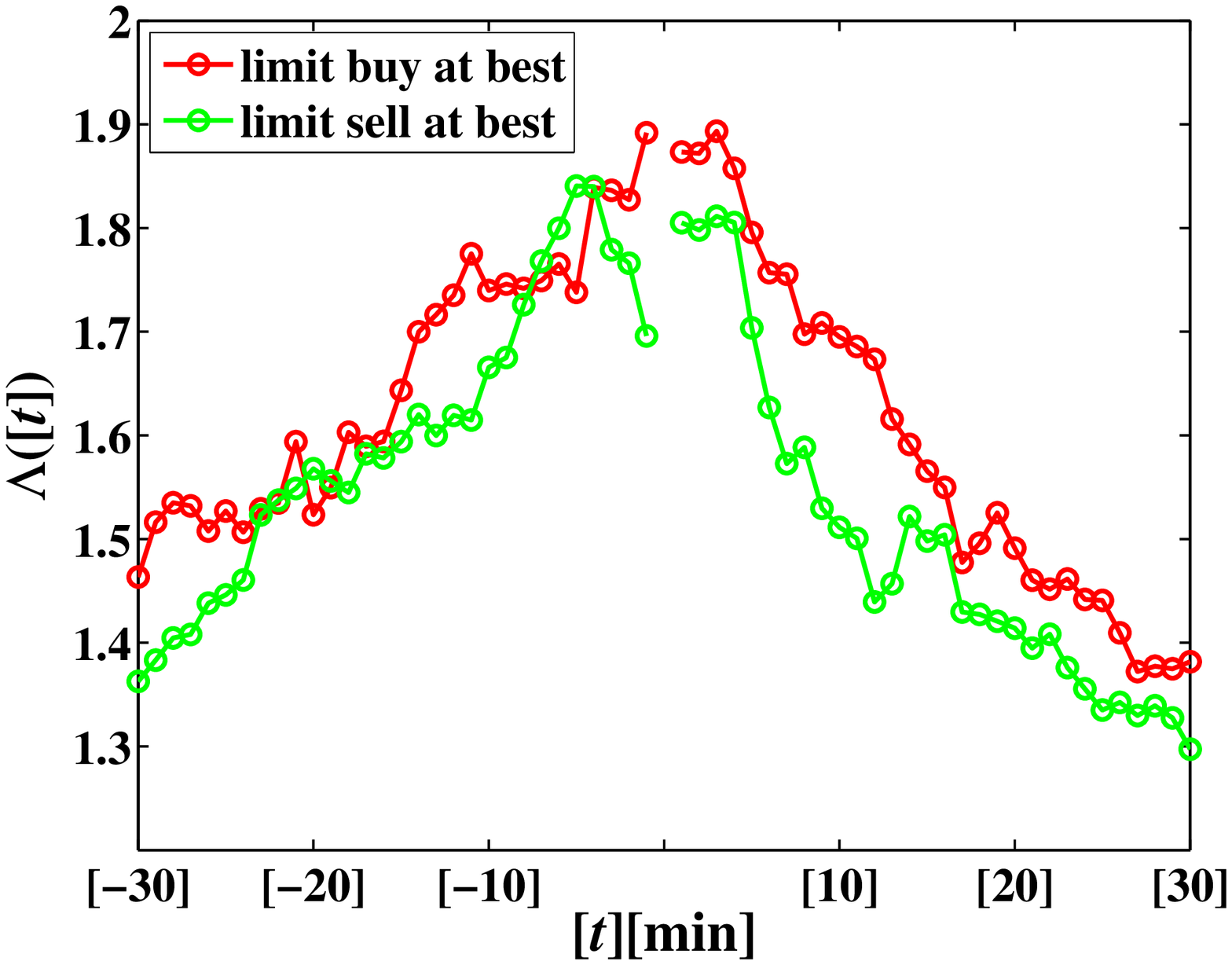}\\
  \includegraphics[width=0.24\linewidth]{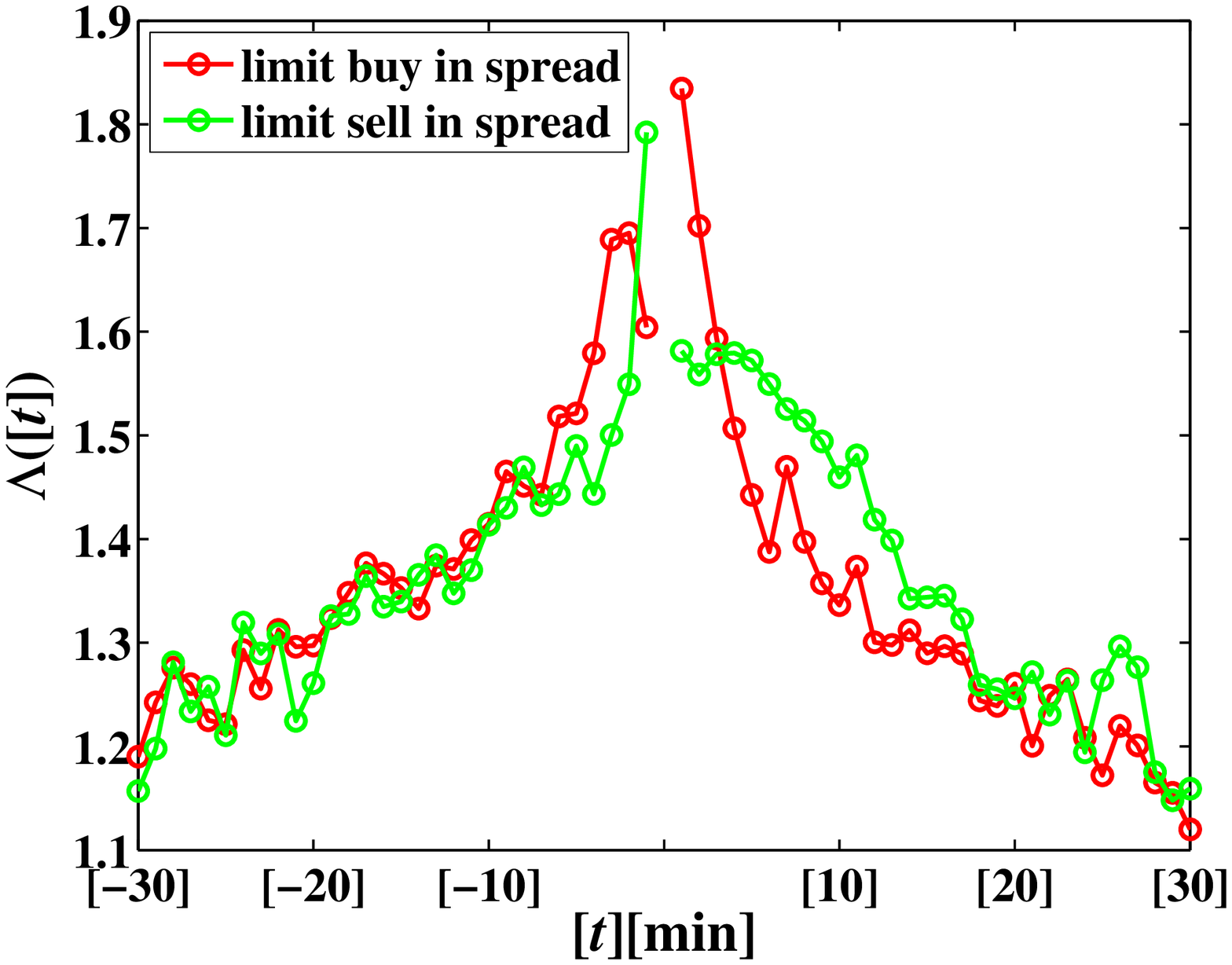}
  \includegraphics[width=0.24\linewidth]{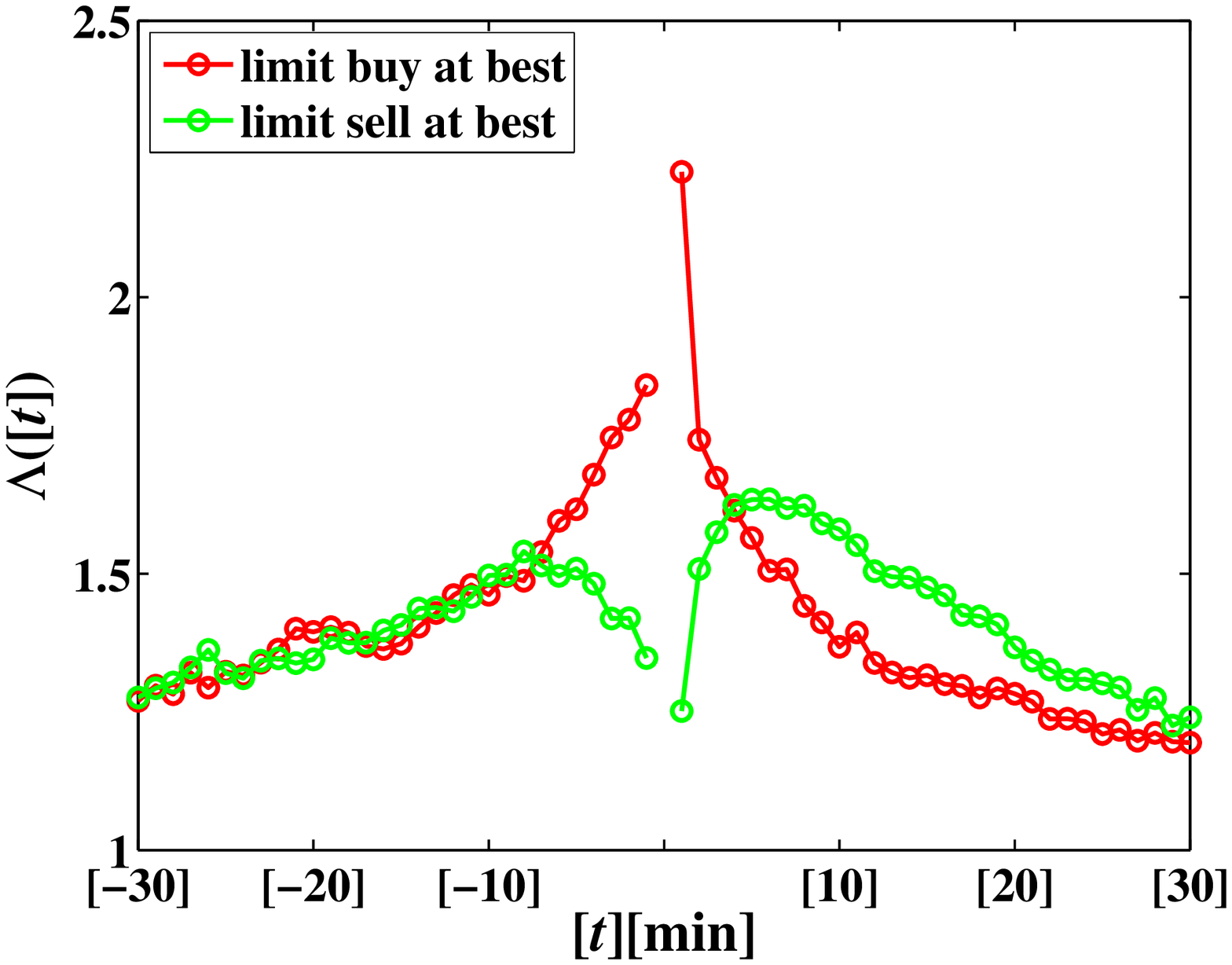}
  \includegraphics[width=0.24\linewidth]{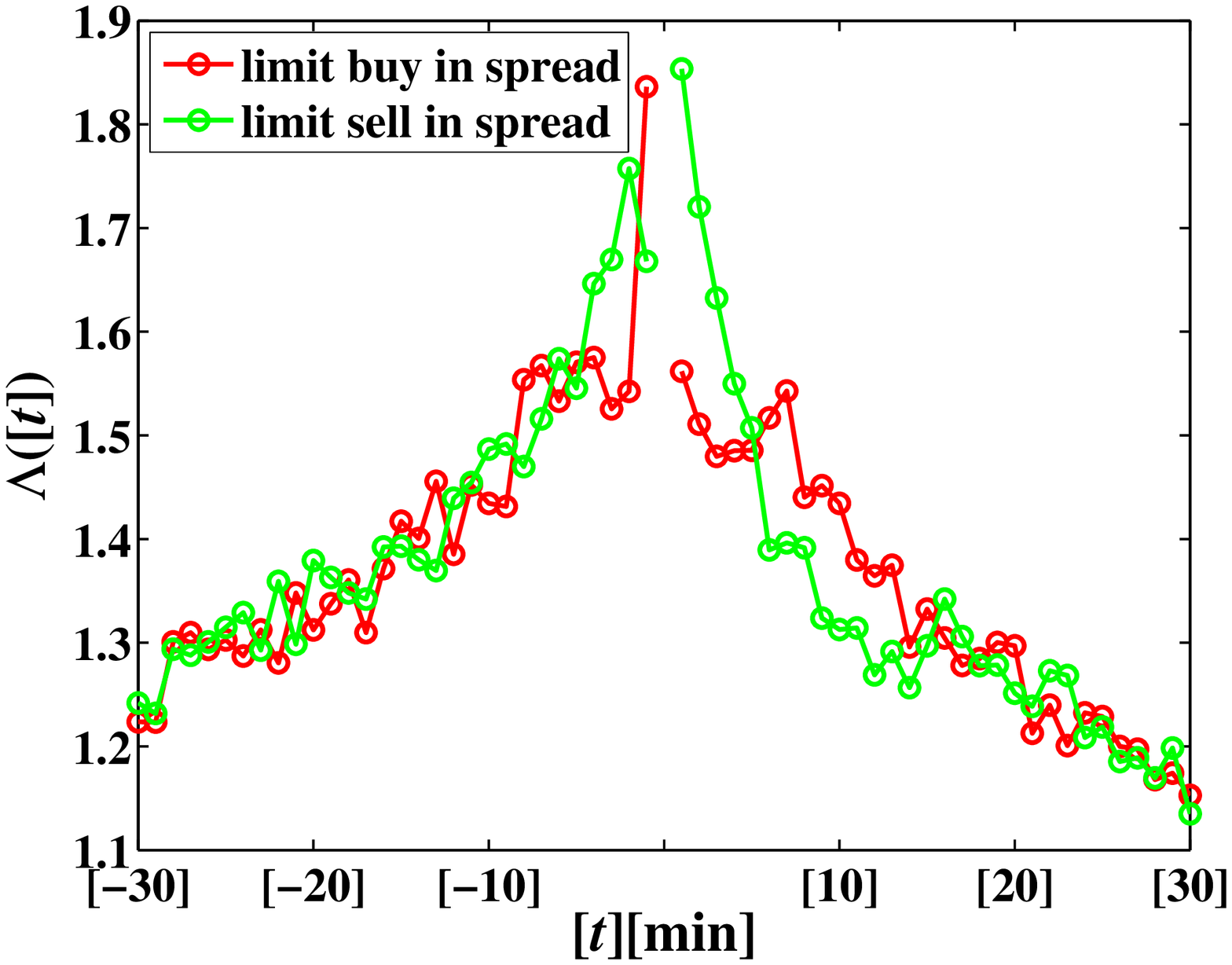}
  \includegraphics[width=0.24\linewidth]{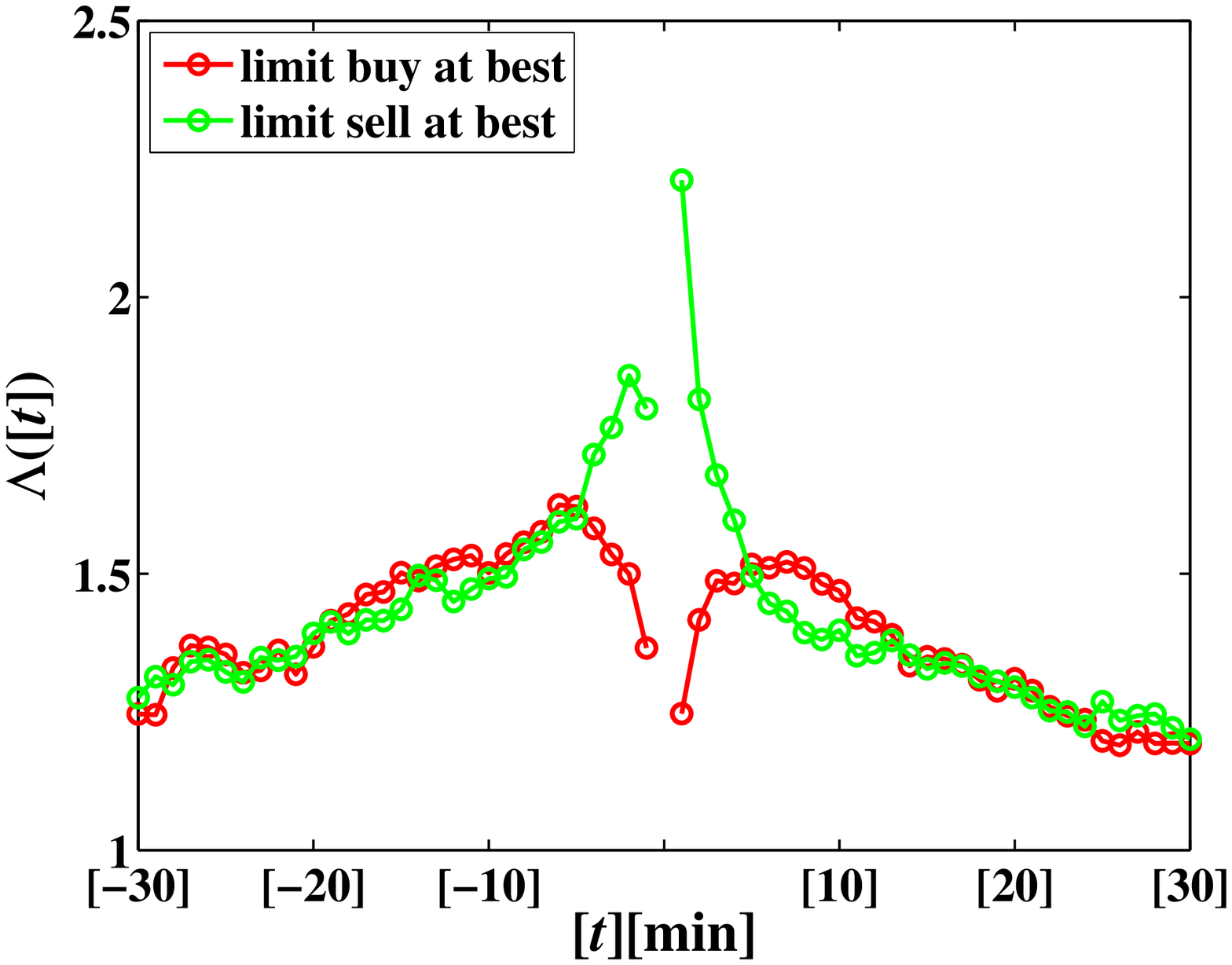}\\
  \includegraphics[width=0.24\linewidth]{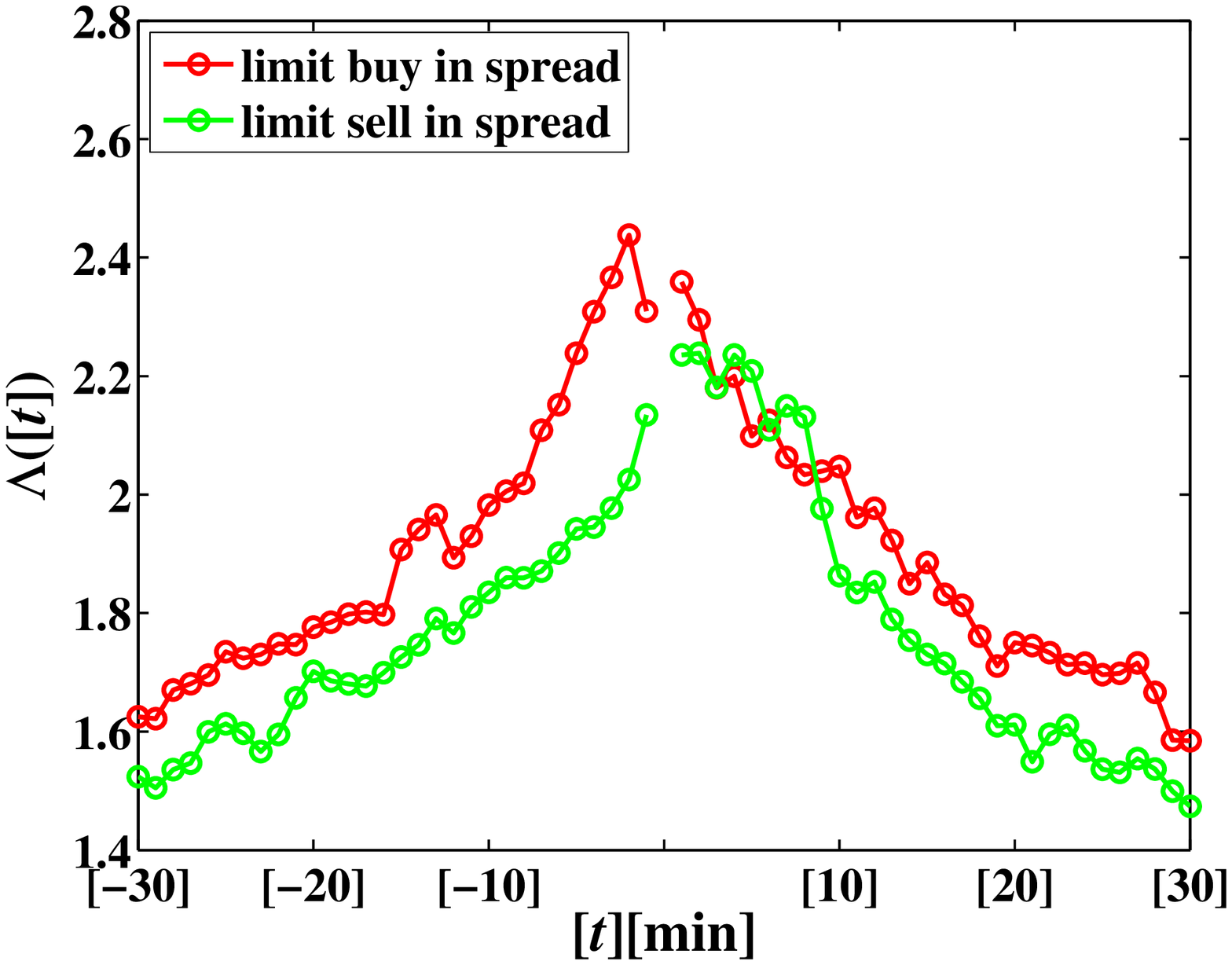}
  \includegraphics[width=0.24\linewidth]{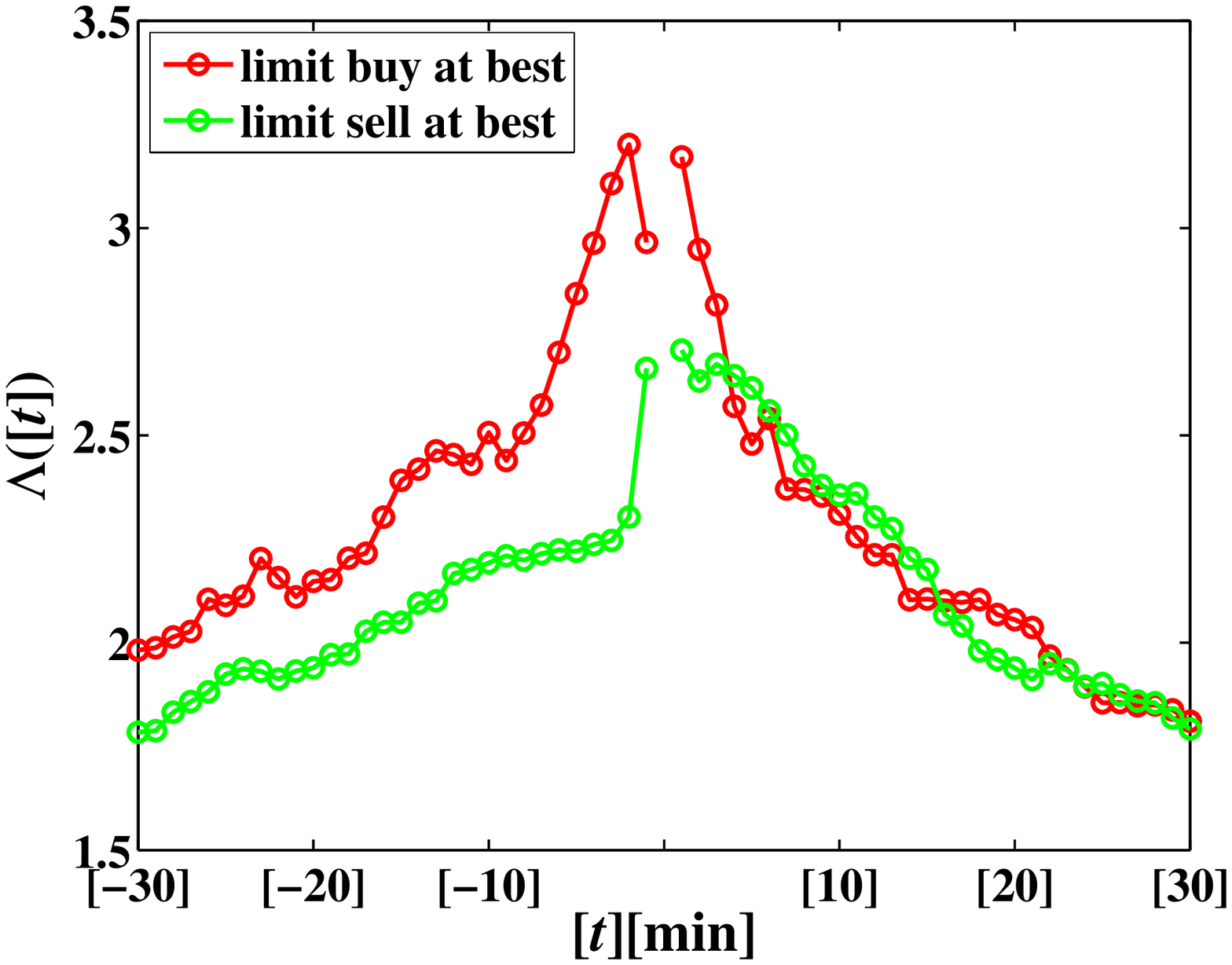}
  \includegraphics[width=0.24\linewidth]{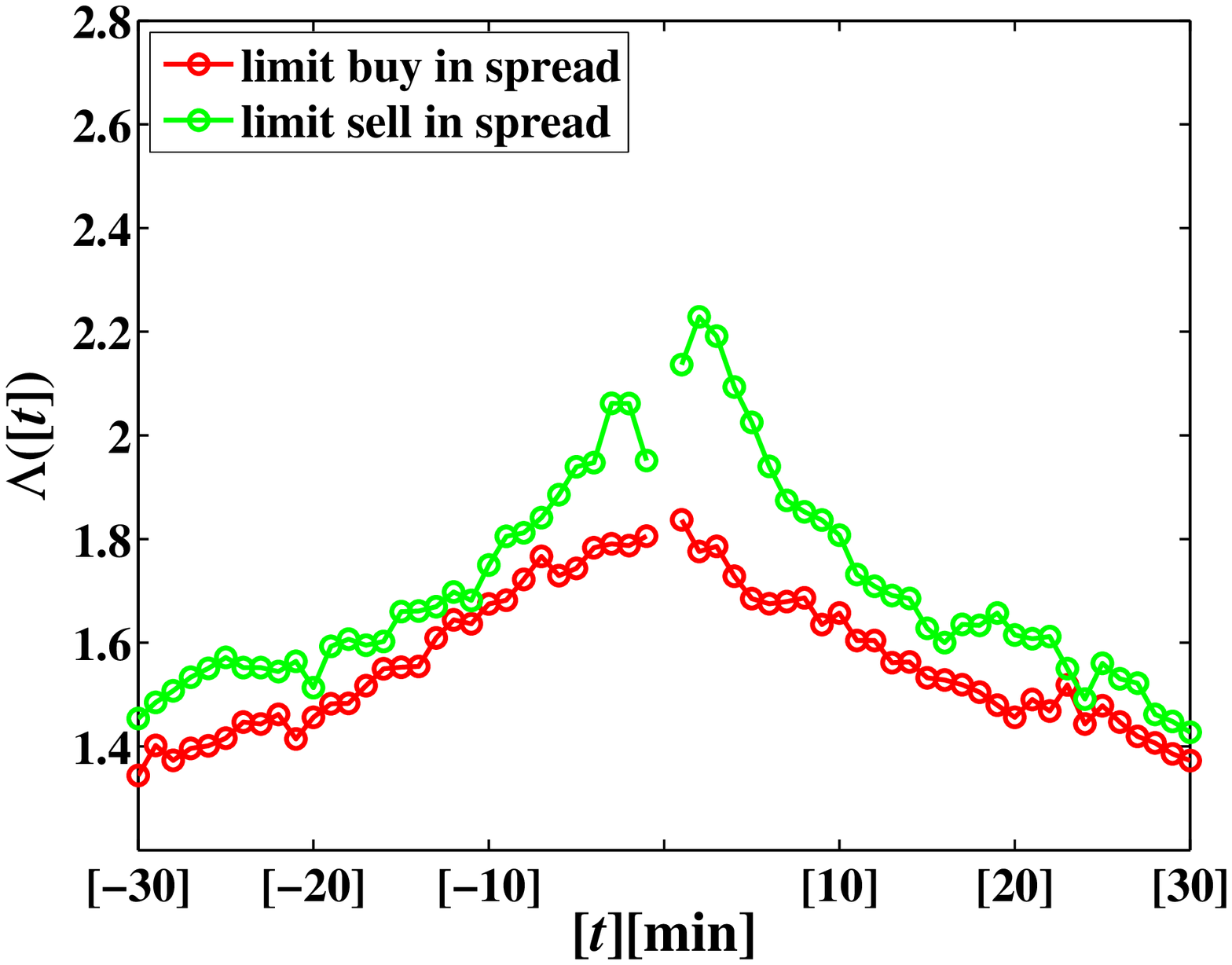}
  \includegraphics[width=0.24\linewidth]{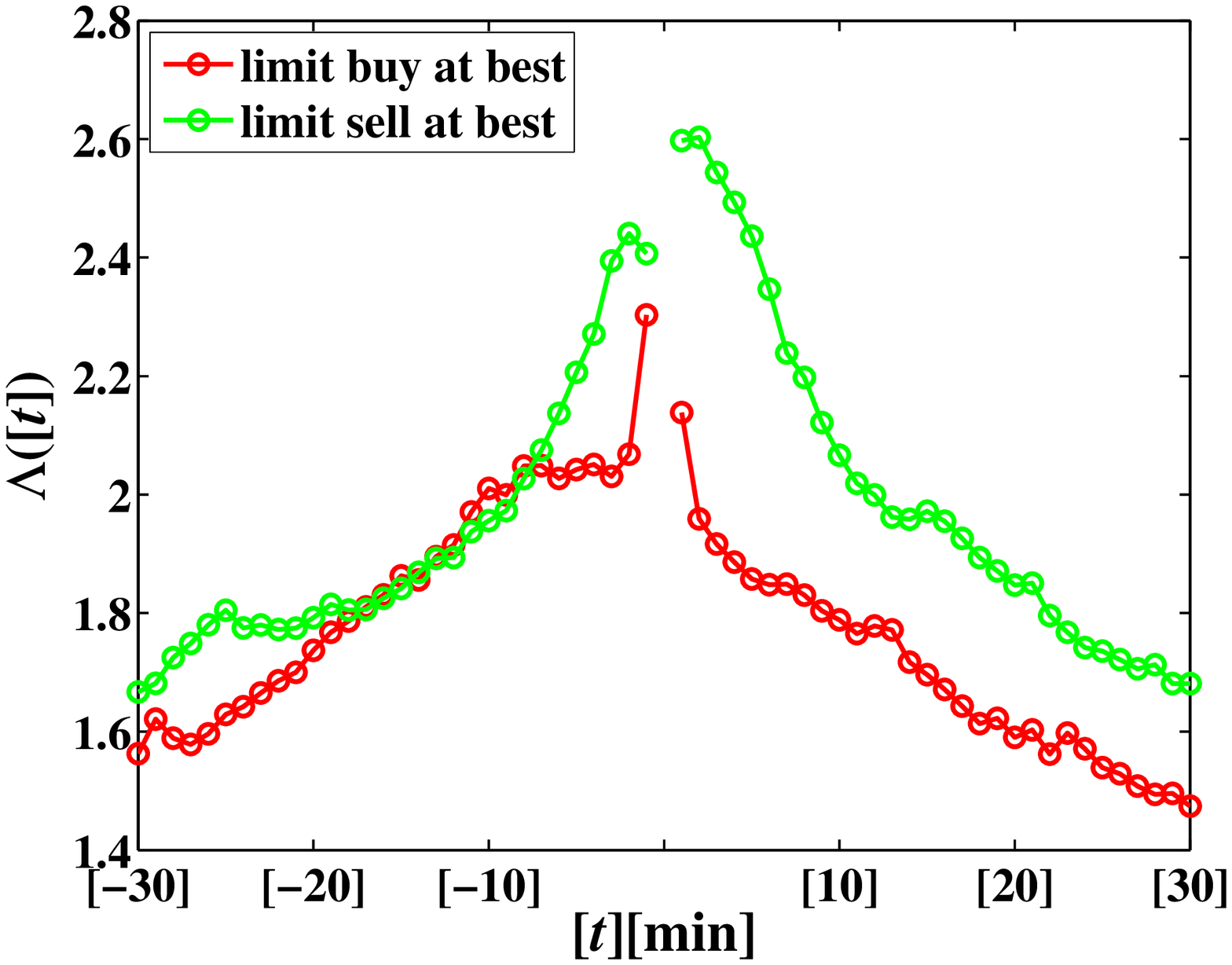}
  \vskip  -0.5800\textwidth   \hskip   -0.985\textwidth (a)
  \vskip  -0.0355\textwidth   \hskip   -0.485\textwidth (b)
  \vskip  -0.0355\textwidth   \hskip   +0.010\textwidth (c)
  \vskip  -0.0355\textwidth   \hskip   +0.505\textwidth (d)
  \vskip  +0.1555\textwidth   \hskip   -0.985\textwidth (e)
  \vskip  -0.0355\textwidth   \hskip   -0.485\textwidth (f)
  \vskip  -0.0355\textwidth   \hskip   +0.010\textwidth (g)
  \vskip  -0.0355\textwidth   \hskip   +0.505\textwidth (h)
  \vskip  +0.1555\textwidth   \hskip   -0.985\textwidth (i)
  \vskip  -0.0355\textwidth   \hskip   -0.485\textwidth (j)
  \vskip  -0.0355\textwidth   \hskip   +0.010\textwidth (k)
  \vskip  -0.0355\textwidth   \hskip   +0.505\textwidth (l)
  \vskip  +0.1555\textwidth
  \caption{\label{Fig:LOBResilency:Intensity} (Color online) Intensity of limit orders for stock 000858 around effective market orders of different types: Type 1 in plots (a) and (b), Type 7 in plots (c) and (d), Type 2 in plots (e) and (f), Type 8 in plots (g) and (h), Type 3 in plots (i) and (j), and Type 9 in plots (k) and (l).}

\end{figure}

The resiliency behavior of bid-ask spread/depth analyzed above can be attributed to the order flow resiliency essentially. For example, is the bid-ask spread resiliency mainly due to the placement of buy limit orders or sell limit orders in the spread? Is the depth resiliency consistent with the intensity of limit orders placed at the best price level? Here we investigate the evolution of limit order intensity around different types of effective market orders to answer these questions. The empirical results for a representative A-share stock 000858 are illustrated in Fig.~\ref{Fig:LOBResilency:Intensity} for the intensity of limit orders placed in the spread (Type-4 and Type-10) and limit orders at the best price (Type-5 and Type-11). The results are very  similar for other A-share and B-share stocks we investigated.


Fig.~\ref{Fig:LOBResilency:Intensity}(a) presents the evolution of intensity of limit orders placed in the spread (Type-4 for buys and Type-10 for sells) around Type-1 buy market orders with the size greater than the outstanding volume on the best ask and the price higher than the best ask, while Fig.~\ref{Fig:LOBResilency:Intensity}(b) presents the evolution of intensity of limit orders placed at the best price (Type-5 for buys and Type-11 for sells) around Type-1 buy market orders. A Type-1 buy market order appears usually when the sell limit order intensity increases rapidly and is remarkably higher than the buy limit order intensity. This observation indicates that the most aggressive liquidity takers enter the market when there are more liquidity providers on the opposite side. After the arrival of a Type-1 buy market order, the bid-ask spread widens \cite{Zhou-2012-NJP}. More limit orders are placed in one or two minutes and the limit order intensity decreases gradually afterwards to its average level within 30 minutes. Moreover, the intensities of Type-10 (and Type-11) sell limit orders are higher than those of Type-4 (and Type-5) buy limit orders  in the after-period of Type-1 orders shock. This means that, Type-10 (and Type-11) sell limit orders contribute more to the resiliency of spread and LOB depth than Type-4 (and Type-5) buy limit orders. Hence the price reversal behavior is dominant after Type-1 buy market orders. In addition, the intensity of Type-4 buy limit orders increases more than Type-10 sell limit orders right after the Type-1 buy market orders, which can be probably viewed as an indicator of herding behaviors among traders. According to Fig.~\ref{Fig:LOBResilency:Intensity}(c) and Fig.~\ref{Fig:LOBResilency:Intensity}(d), the behaviors around Type-7 sell market orders are similar.

In Fig.~\ref{Fig:LOBResilency:Intensity}(e) and Fig.~\ref{Fig:LOBResilency:Intensity}(f), we present respectively the evolution of intensity of limit orders placed in the spread (Type-4 for buys and Type-10 for sells) and at the best price (Type-5 for buys and Type-11 for sells) around Type-2 buy market orders with the size greater than the outstanding volume on the best ask and the price higher than the best ask. Both intensities of Type-4 and Type-5 buy limit orders increase before Type-2 buy market orders. They continue increasing in the first minute after Type-2 buy market orders and then decay to the average level within about 30 minutes, which implies a herding behavior of buyers, similar as Type-4 and Type-5 buy limit orders after Type-1 buy market orders shown in Fig.~\ref{Fig:LOBResilency:Intensity}(a). The patterns of sell limit orders around Type-2 buy market orders are very different. Before Type-2 buy market orders, the intensity of Type-10 sell limit orders increases, while the intensity of Type-11 sell limit orders increases first and starts to decrease about 10 minutes before Type-2 buy market orders. After Type-2 buy market orders, the intensity of Type-10 sell limit orders that are placed in the spread decreases continuously to the average level within 30 minutes, while the intensity of Type-11 sell limit orders submitted at the best ask decreases immediately in the first minute and then increases in the subsequent a few minutes before decaying to the average level. This phenomenon indicates that buy limit orders contribute more to the resiliency of spread than sell limit orders after Type-2 buy limit orders and the price continues to rise. It is also worth noting that, comparing the intensity of limit orders at at best quotes shown in Fig.~\ref{Fig:LOBResilency:Intensity}(f) and Fig.~\ref{Fig:LOBResilency:Intensity}(h) with the depth at the best quotes (column 3 and column 4 in Fig.~\ref{Fig:LOBResilency:Depth:Type}) after Type-2 and Type-8 market orders, the limit order intensity play an obvious effect on the bid-ask depth balance since more limit orders are placed to the side with lower depth. The patterns of limit order intensity around Type-8 sell market orders in Fig.~\ref{Fig:LOBResilency:Intensity}(g) and Fig.~\ref{Fig:LOBResilency:Intensity}(h) are similar.

In Fig.~\ref{Fig:LOBResilency:Intensity}(i) and Fig.~\ref{Fig:LOBResilency:Intensity}(j), we illustrate respectively the evolution of intensity of limit orders placed in the spread (Type-4 for buys and Type-10 for sells) and at the best price (Type-5 for buys and Type-11 for sells) around Type-3 buy market orders with the size less than the outstanding volume on the best ask and the price not less than the best ask. Before Type-3 buy market orders, the intensities of buy and sell limit orders increase continuously and buy limit orders have higher intensity than sell limit orders. After the entering of Type-3 buy market orders, the intensities of different types of limit orders increase slightly and then decay to the average level within 30 minutes. In the first a few minute after Type-3 buy market orders, buy limit orders have higher intensity than sell limit orders, indicating again that buy limit orders contribute more to the resiliency of spread than sell limit orders and the price continues to rise. The observations for limit order intensity around Type-9 sell market orders in Fig.~\ref{Fig:LOBResilency:Intensity}(k) and Fig.~\ref{Fig:LOBResilency:Intensity}(l) are qualitatively similar. However, the intensity difference between sell limit orders and buy limit orders after Type-9 sell market orders in Fig.~\ref{Fig:LOBResilency:Intensity}(k) and Fig.~\ref{Fig:LOBResilency:Intensity}(l) is larger than the intensity difference between buy limit orders and sell limit orders after Type-3 buy market orders in Fig.~\ref{Fig:LOBResilency:Intensity}(i) and Fig.~\ref{Fig:LOBResilency:Intensity}(j), suggesting that traders are more sensitive to bad news and have stronger herding behaviors.



\begin{figure}[!htb]
  \centering
  \includegraphics[width=0.24\linewidth]{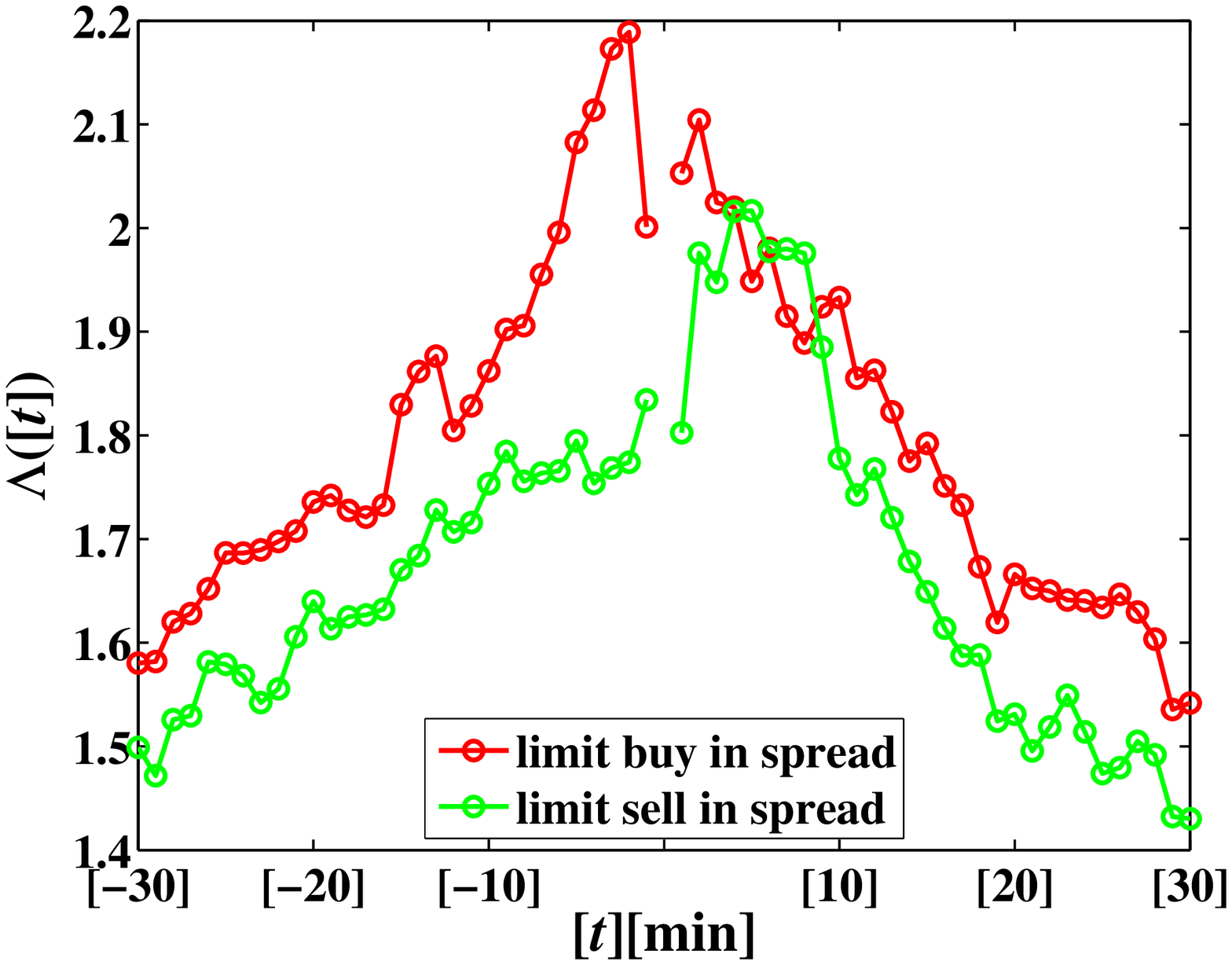}
  \includegraphics[width=0.24\linewidth]{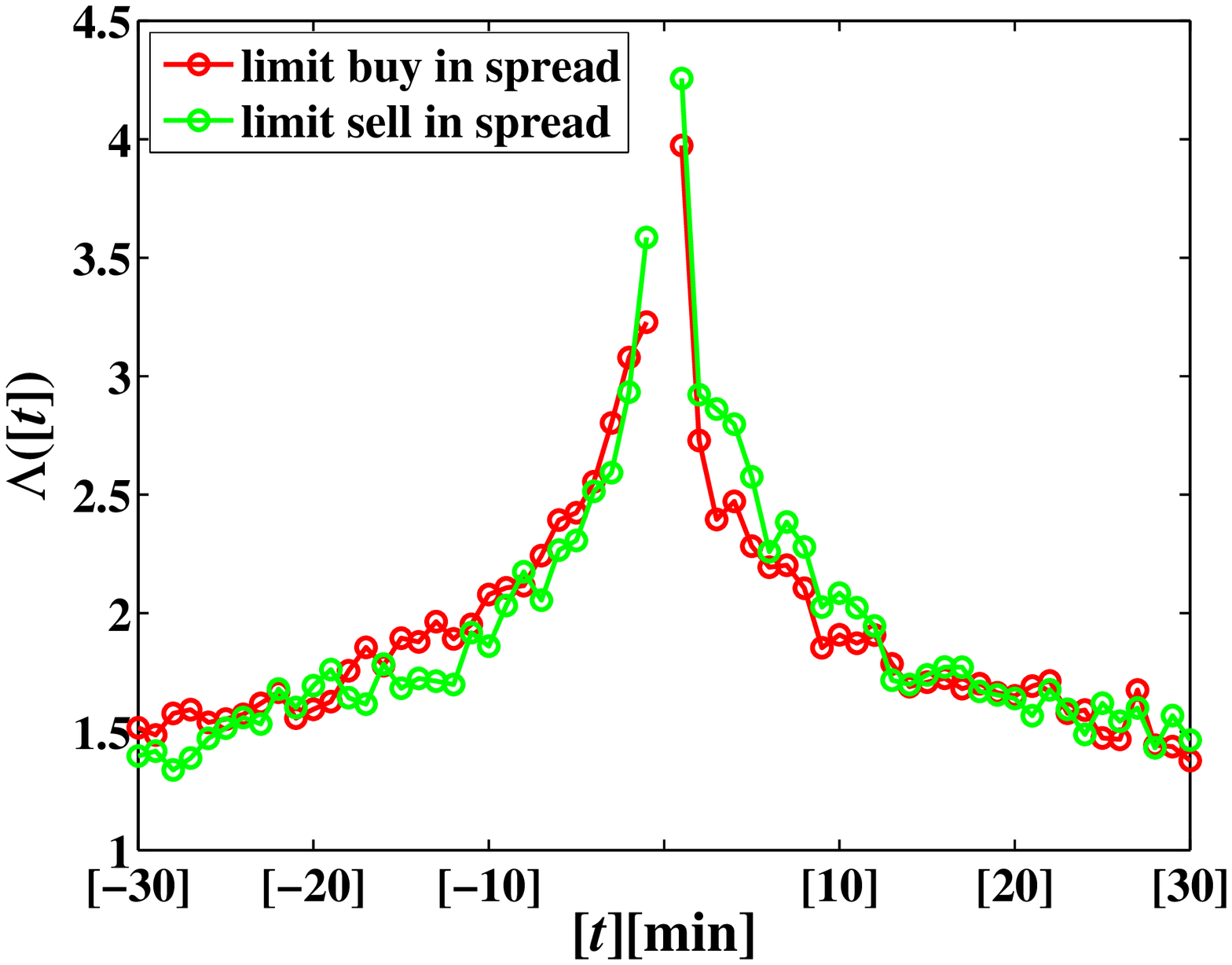}
  \includegraphics[width=0.24\linewidth]{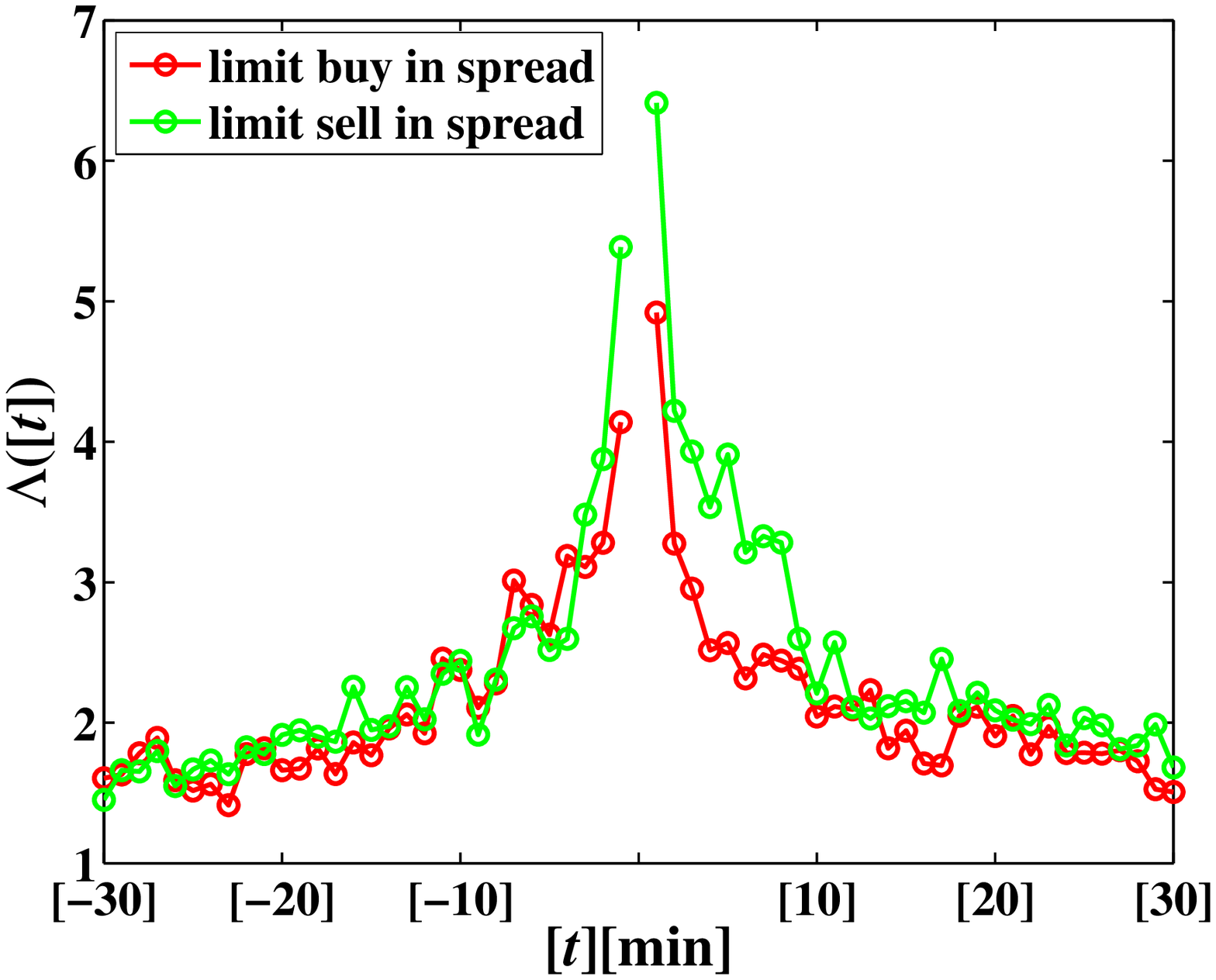}
  \includegraphics[width=0.24\linewidth]{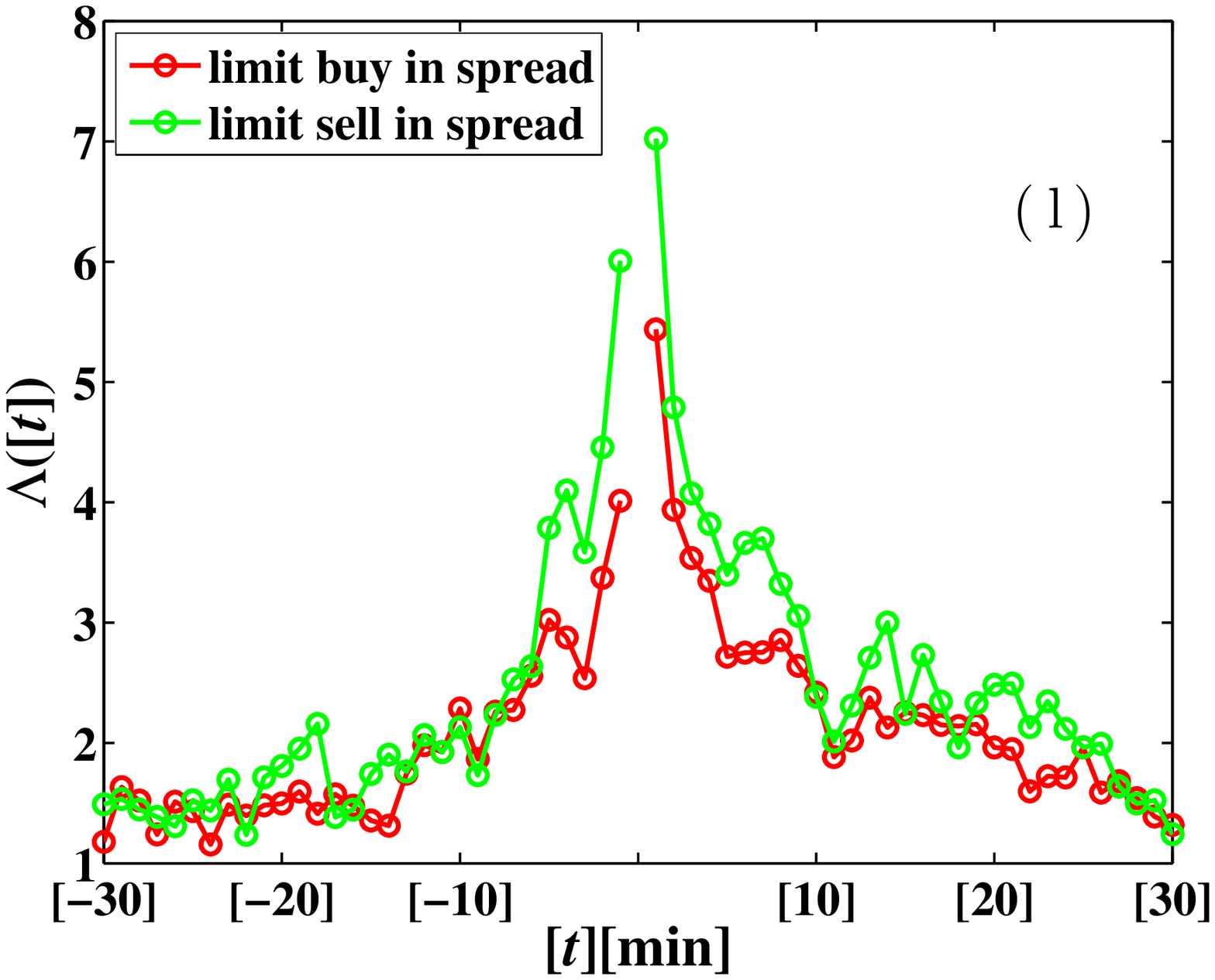}\\
  \includegraphics[width=0.24\linewidth]{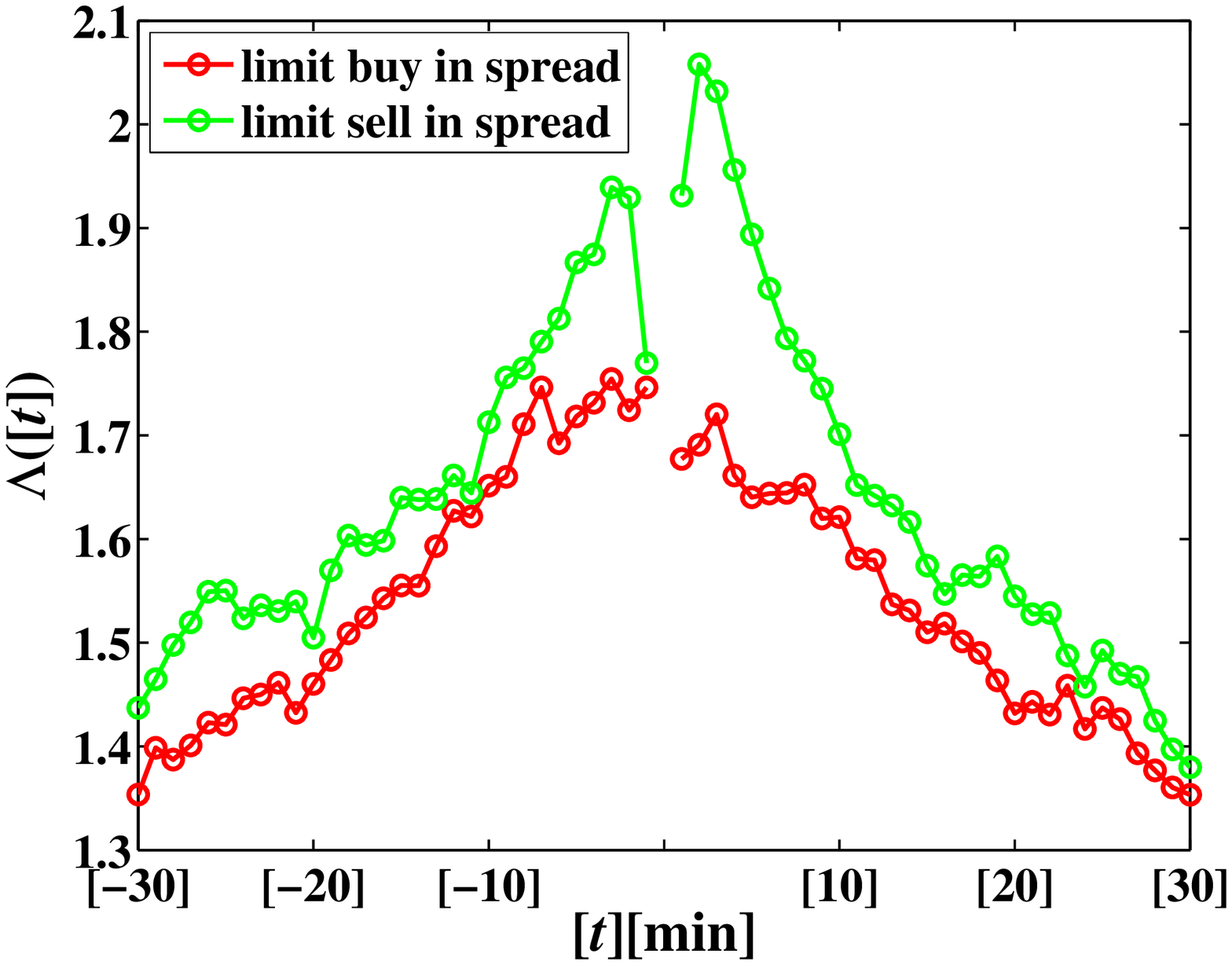}
  \includegraphics[width=0.24\linewidth]{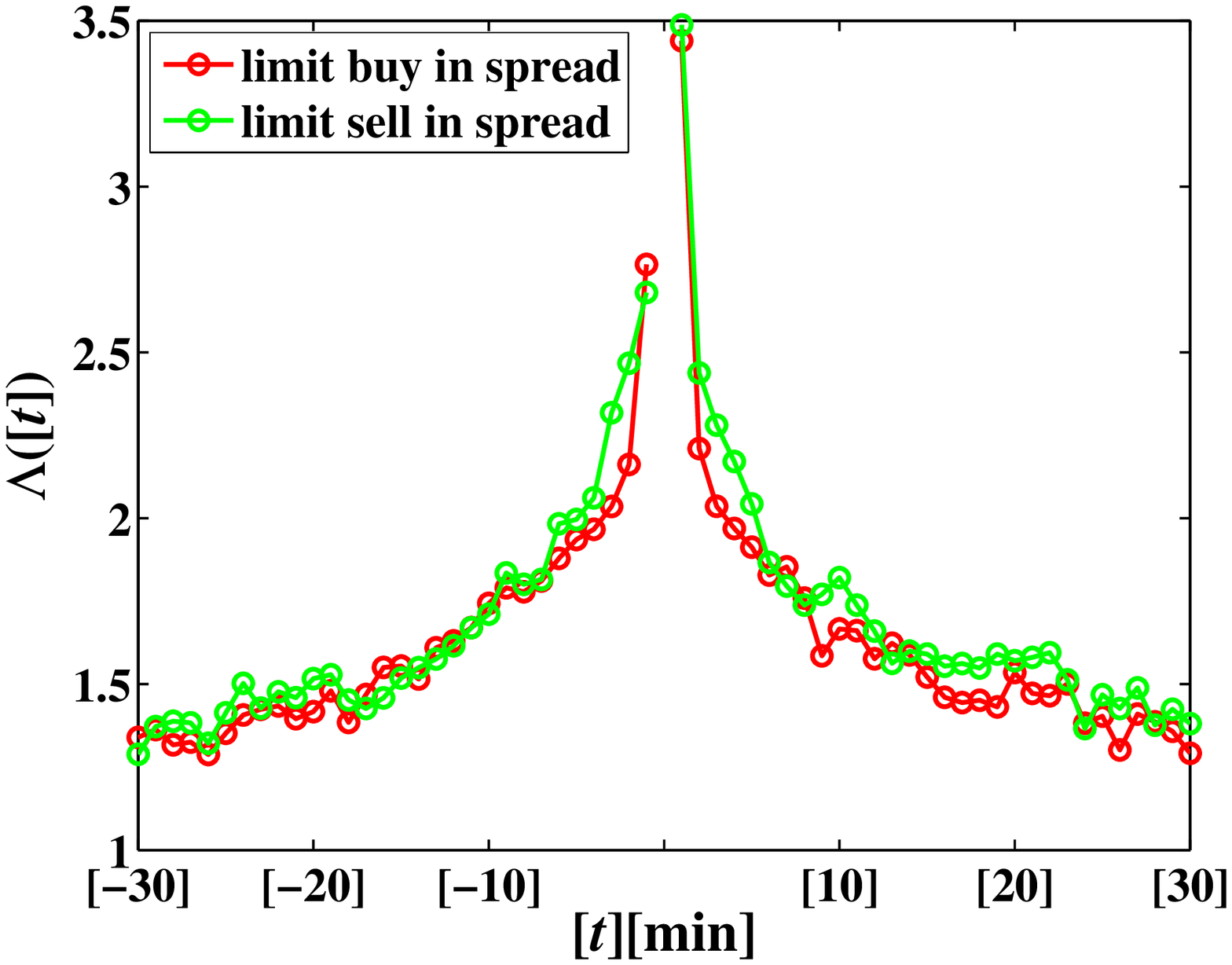}
  \includegraphics[width=0.24\linewidth]{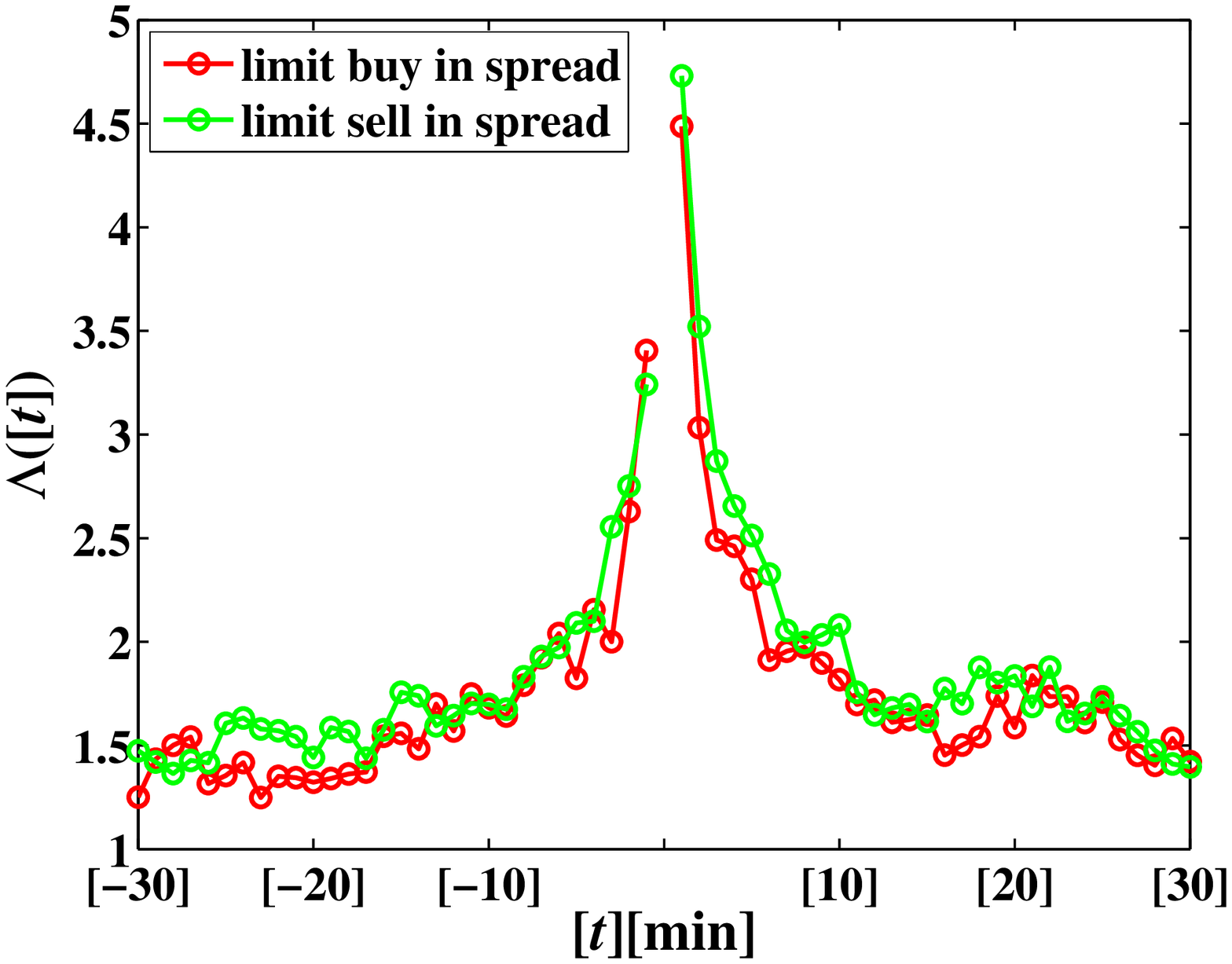}
  \includegraphics[width=0.24\linewidth]{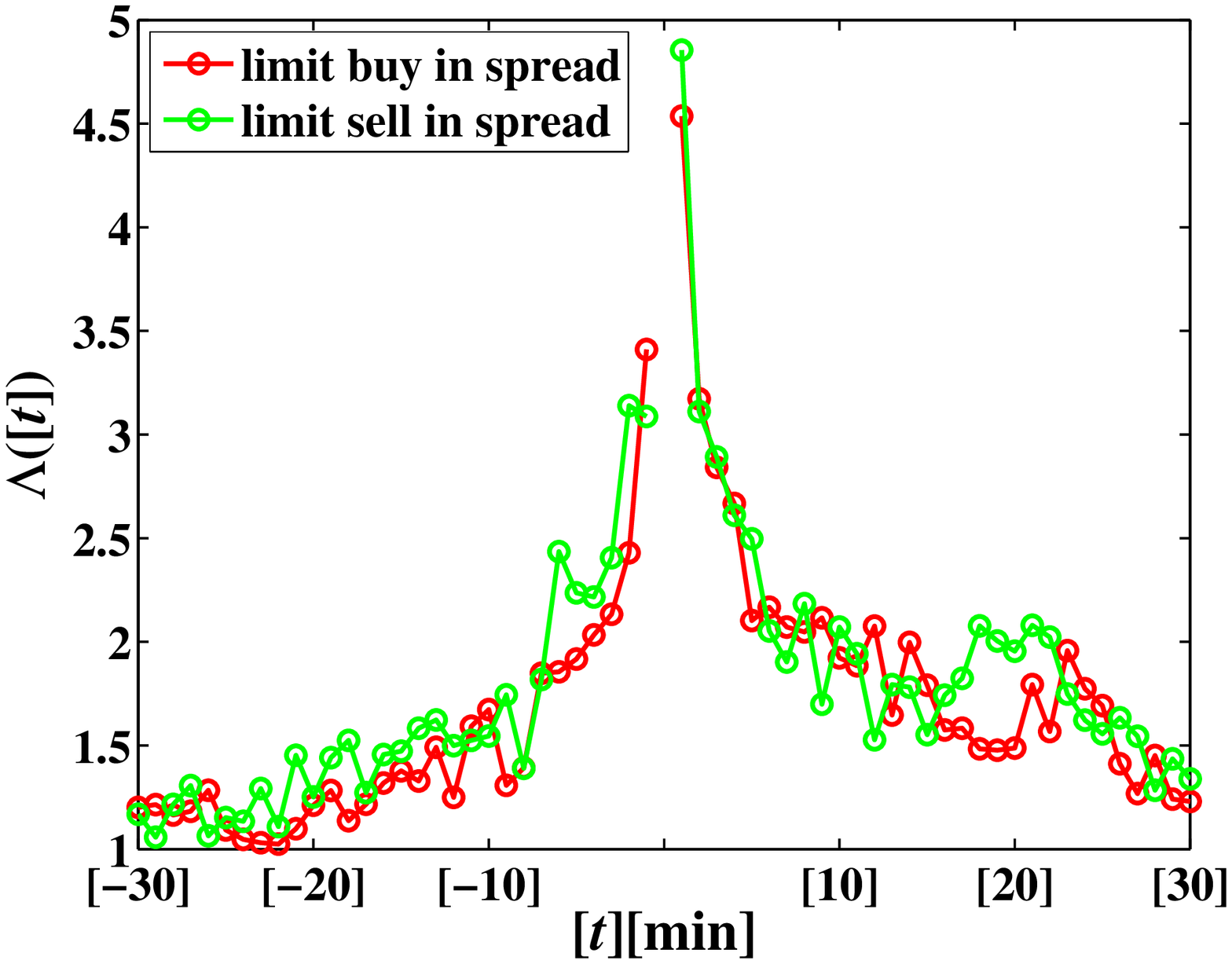}
  \includegraphics[width=0.24\linewidth]{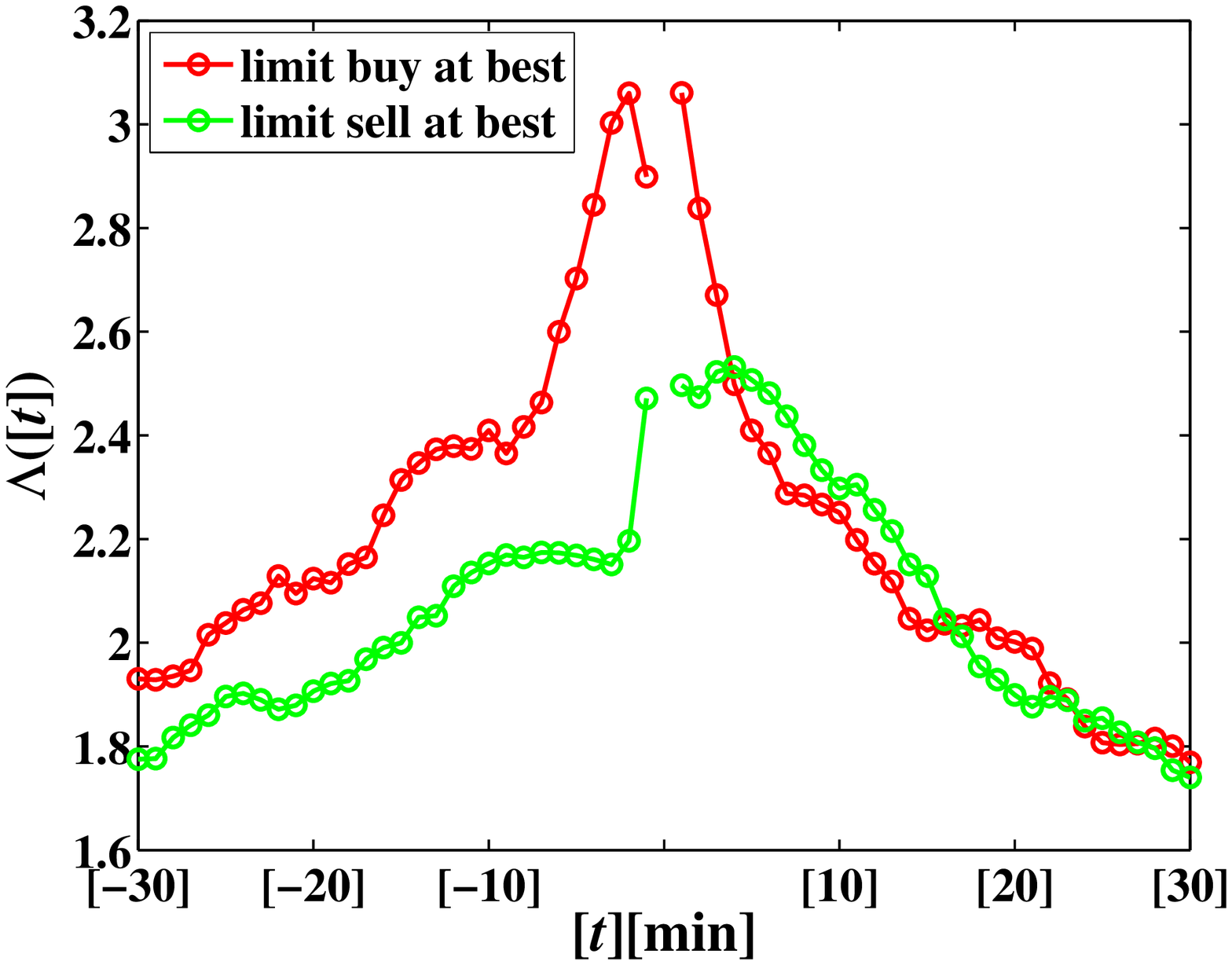}
  \includegraphics[width=0.24\linewidth]{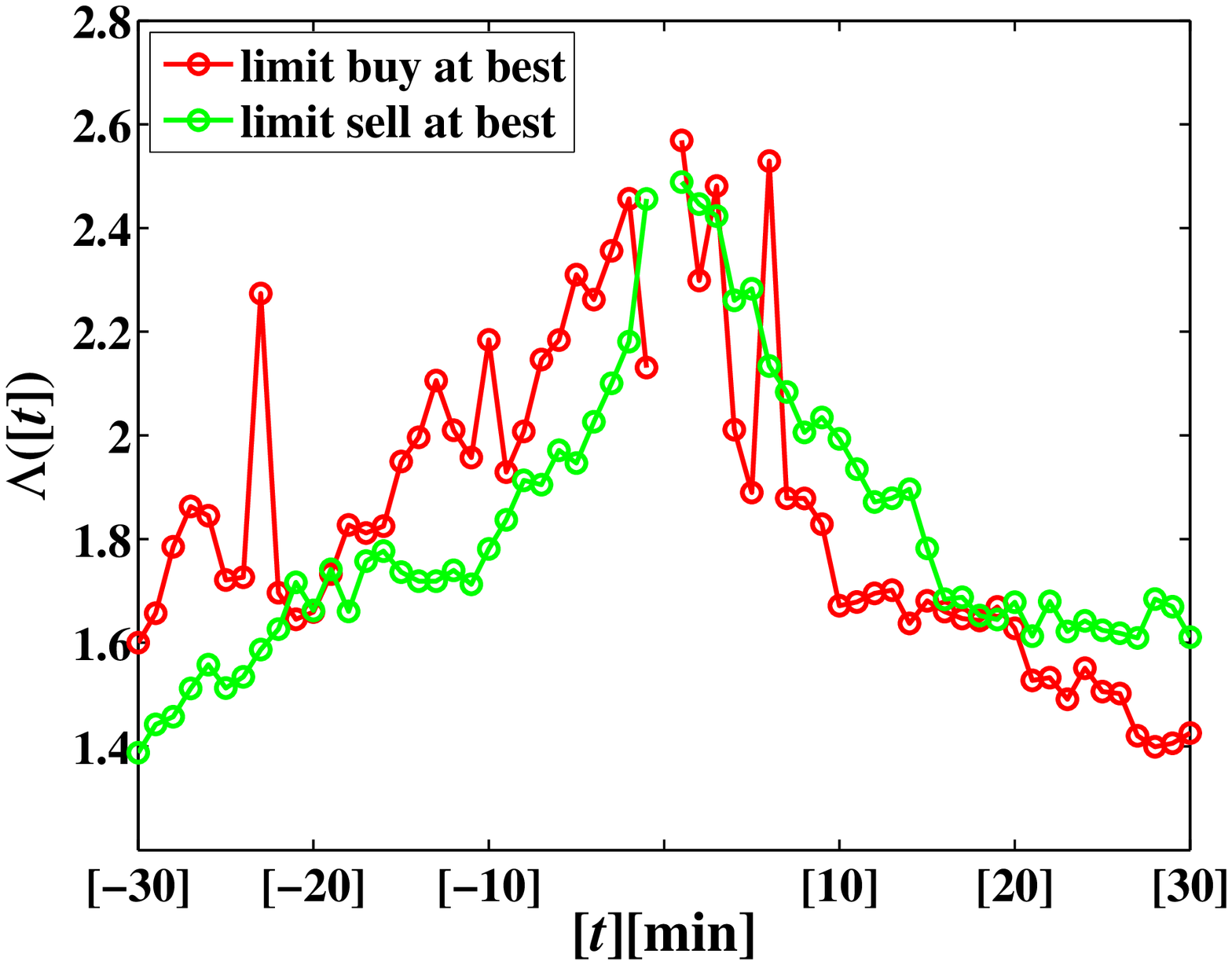}
  \includegraphics[width=0.24\linewidth]{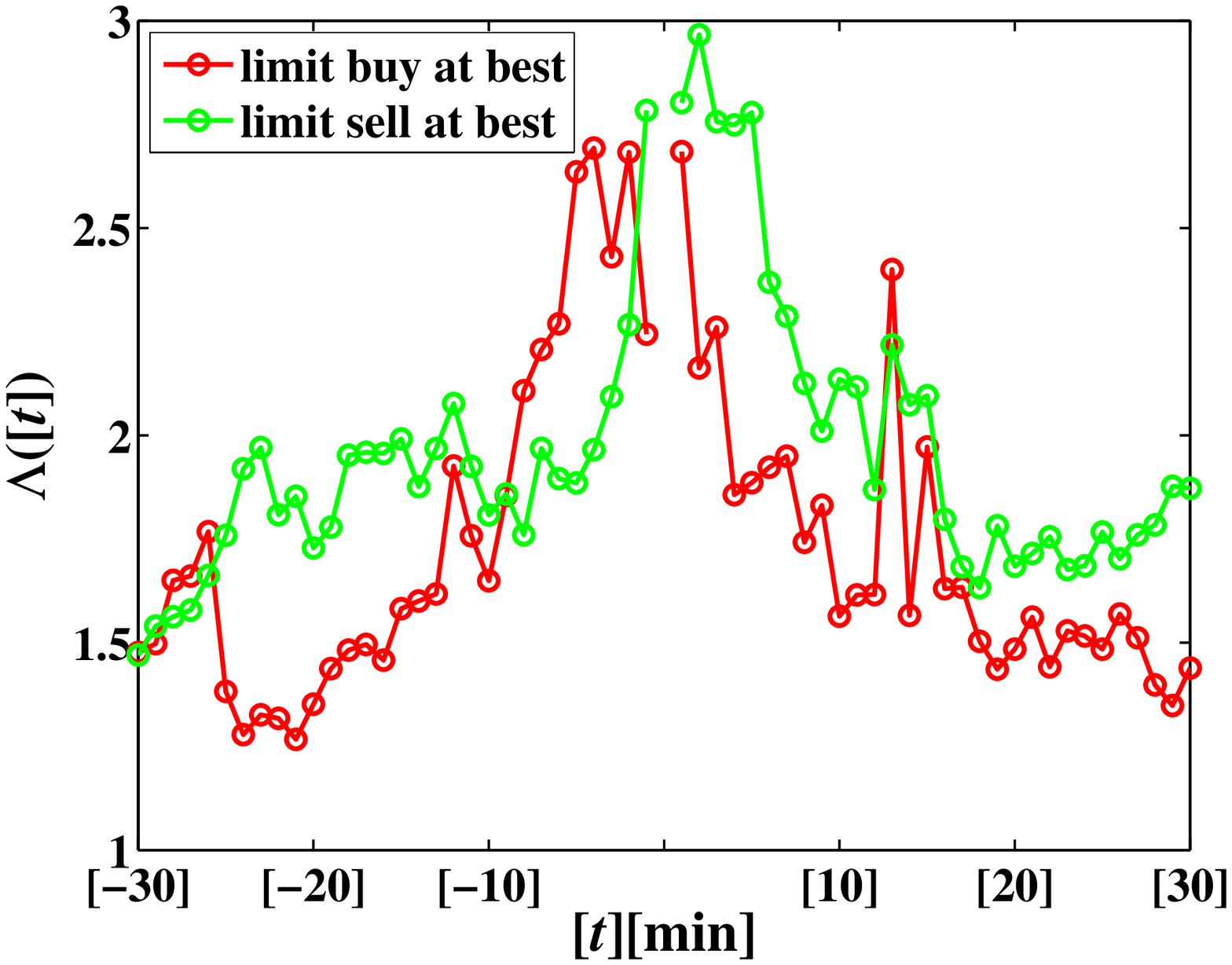}
  \includegraphics[width=0.24\linewidth]{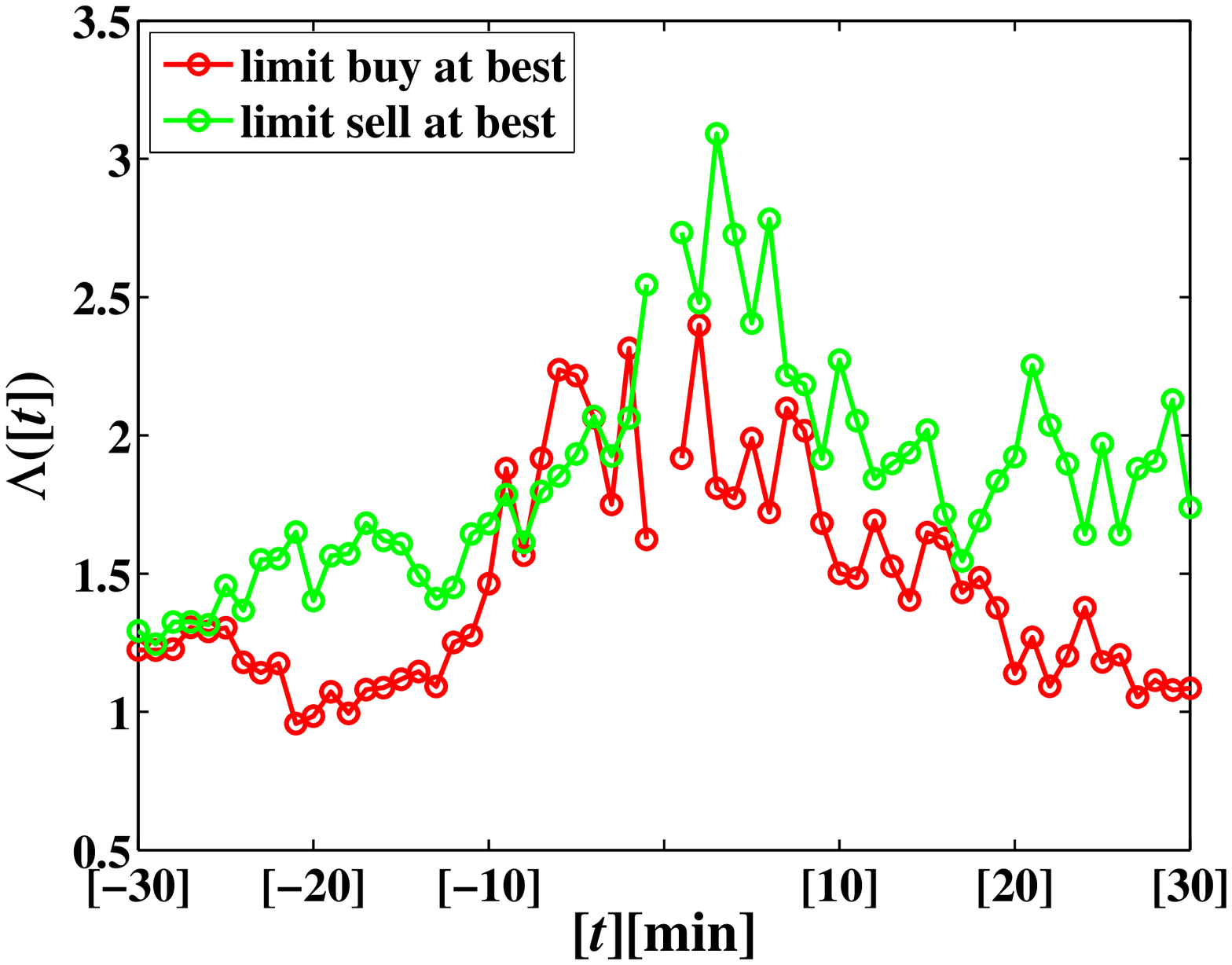}\\
  \includegraphics[width=0.24\linewidth]{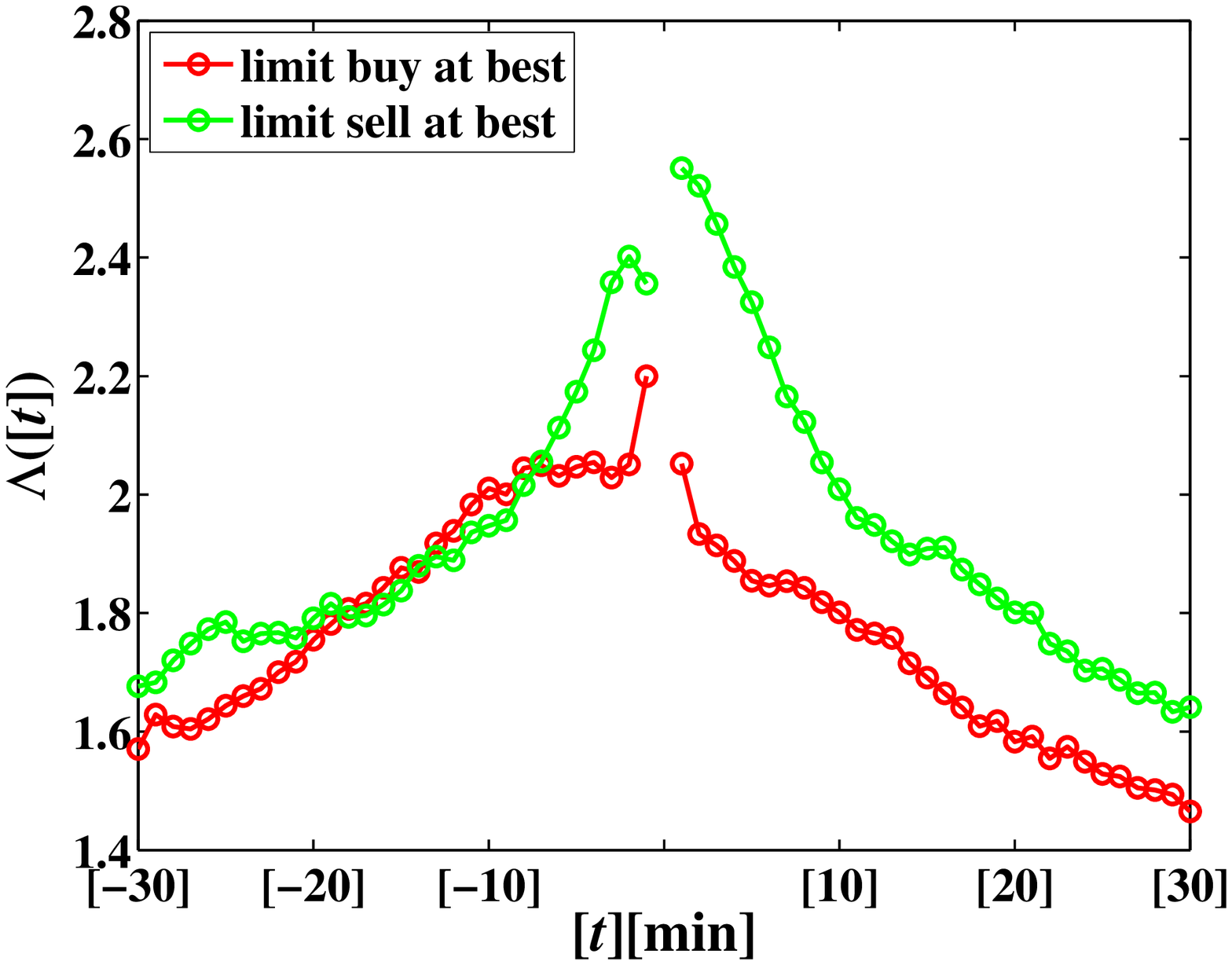}
  \includegraphics[width=0.24\linewidth]{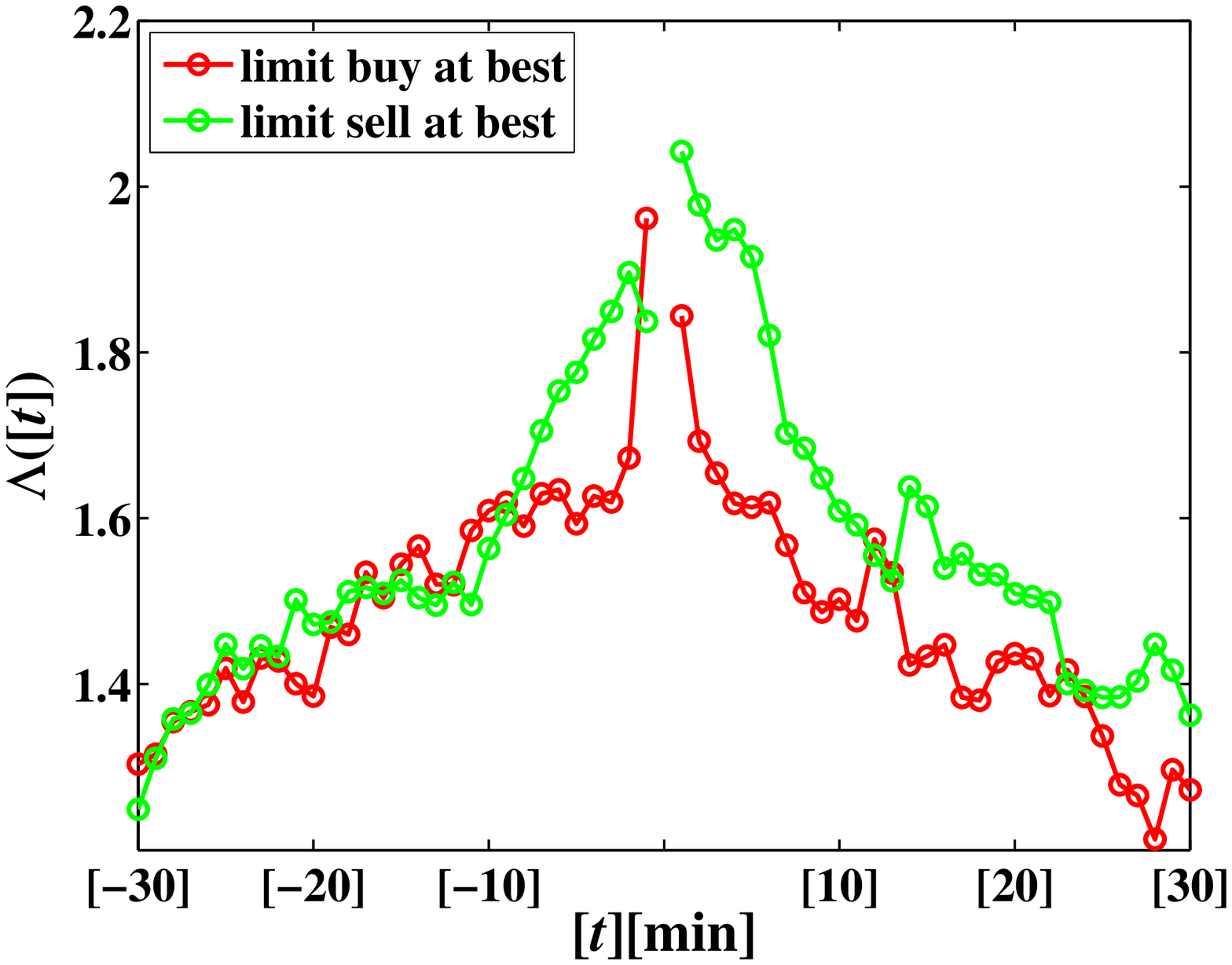}
  \includegraphics[width=0.24\linewidth]{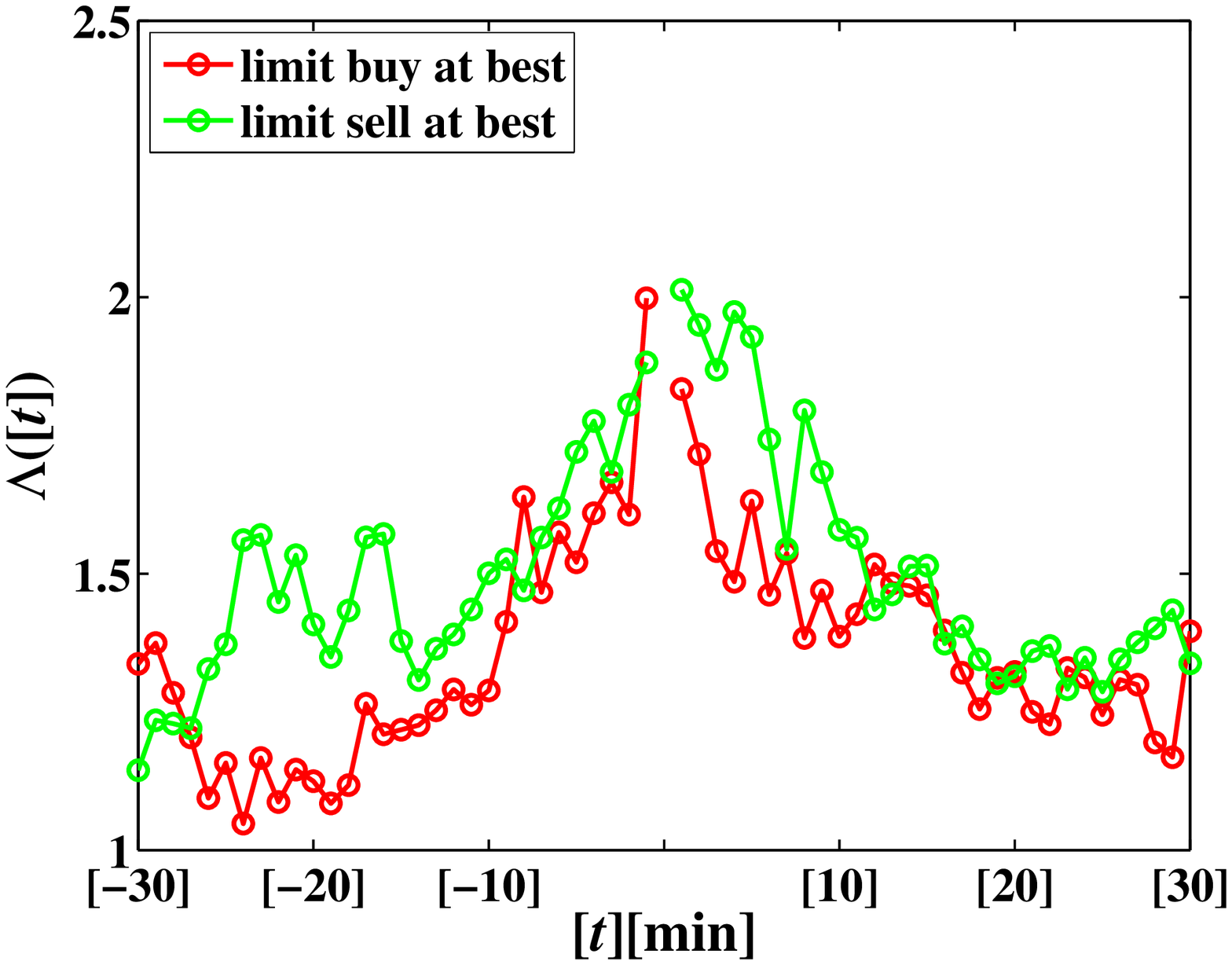}
  \includegraphics[width=0.24\linewidth]{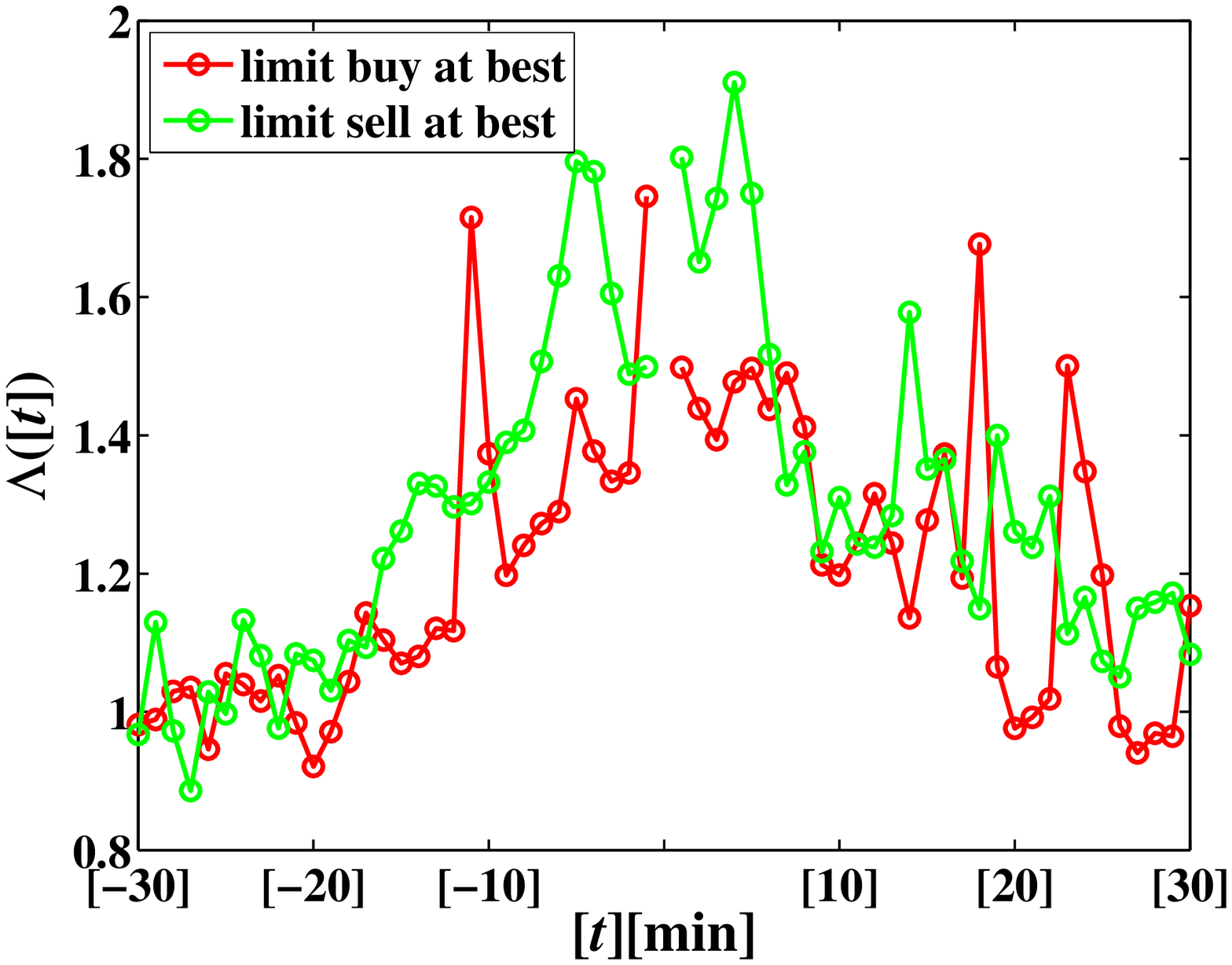}
  \vskip  -0.7660\textwidth   \hskip   -0.985\textwidth (a)
  \vskip  -0.0356\textwidth   \hskip   -0.485\textwidth (b)
  \vskip  -0.0356\textwidth   \hskip   +0.010\textwidth (c)
  \vskip  -0.0356\textwidth   \hskip   +0.505\textwidth (d)
  \vskip  +0.1500\textwidth   \hskip   -0.985\textwidth (e)
  \vskip  -0.0356\textwidth   \hskip   -0.485\textwidth (f)
  \vskip  -0.0356\textwidth   \hskip   +0.010\textwidth (g)
  \vskip  -0.0356\textwidth   \hskip   +0.505\textwidth (h)
  \vskip  +0.545\textwidth
  \vskip  -0.390\textwidth   \hskip   -0.985\textwidth (i)
  \vskip  -0.0356\textwidth   \hskip   -0.480\textwidth (j)
  \vskip  -0.0356\textwidth   \hskip   +0.010\textwidth (k)
  \vskip  -0.0356\textwidth   \hskip   +0.505\textwidth (l)
  \vskip  +0.150\textwidth   \hskip   -0.985\textwidth (m)
  \vskip  -0.0356\textwidth   \hskip   -0.485\textwidth (n)
  \vskip  -0.0356\textwidth   \hskip   +0.010\textwidth (o)
  \vskip  -0.0356\textwidth   \hskip   +0.505\textwidth (p)
  \vskip  +0.155\textwidth  \caption{\label{Fig:LOBResilency:Intensity:DifferentSpread} (Color online) Impact of initial bid-ask spread on the intensity of limit orders around effective market orders for stock 000858. The initial spreads for the four columns are $s(0^-)=0.01$, $s(0^-)=0.02$, $s(0^-)=0.03$ and $s(0^-)\geq0.04$, respectively. (a-d) Intensity of limit orders in the spread around effective buy market orders. (e-h) Intensity of limit orders in the spread around effective sell market orders. (i-l) Intensity of limit orders at the best quotes around effective buy market orders. (m-p) Intensity of limit orders at the best quotes around effective sell market orders. Time $t=0$ corresponds to the transaction caused by the effective market order.}
\end{figure}

Plots (a-h) of Fig.~\ref{Fig:LOBResilency:Intensity:DifferentSpread} show the evolution of intensity of limit orders in the spread (Type-4 and Type-10 orders) around effective market orders with different initial spreads. Similar with that of spread and depth, the evolutionary patterns of limit orders intensity when $s(0^-)=0.01$ are significantly different from other cases. The intensity curves of buy limit orders and sell limit orders basically overlap for $s(0^-)\geq0.02$ in Fig.~\ref{Fig:LOBResilency:Intensity:DifferentSpread}(b-d) and Fig.~\ref{Fig:LOBResilency:Intensity:DifferentSpread}(f-h). This indicates that no matter the effective market order shock is buy or sell, the intensity of the following buy limit orders has no obvious difference from that of following sell limit orders for the case of $s(0^-)\geq0.02$. Both of them show higher intensities within one minute after the shock than their respective parts within one minute before the shock, and then these intensities of limit orders slowly decay to their normal levels within 30 minutes. This means that the effective market orders produce \emph{symmetrical} stimulus to Type-4 orders and Type-10 orders under the case of $s(0^-)\geq0.02$. However, for the case of $s(0^-)=0.01$, the intensity curve of buy limit orders does not overlap with that of sell limit orders. After effective buy market orders, the intensity of Type-4 buy limit orders is higher than or comparable to the intensity of Type-10 sell limit orders in Fig.~\ref{Fig:LOBResilency:Intensity:DifferentSpread}(a). Accordingly, effective sell market orders are followed by more Type-10 sell limit orders than Type-4 buy limit orders, as shown in Fig.~\ref{Fig:LOBResilency:Intensity:DifferentSpread}(e). We find that the corresponding intensity curves in Fig.~\ref{Fig:LOBResilency:Intensity:DifferentSpread}(a,e) and in Fig.~\ref{Fig:LOBResilency:Intensity}(i,k) are quite similar, indicating that the price continuation behavior after Type-3 and Type-9 effective market orders is especially significant when the initial spread is 0.01 CNY.

Similar analysis can be performed on the limit orders placed at the best quotes (Type-5 and Type-11 orders). Plots (i-p) of Fig.~\ref{Fig:LOBResilency:Intensity:DifferentSpread} show the intensity of limit orders at the best quotes around effective market orders with different initial spreads. Similar with Fig.~\ref{Fig:LOBResilency:Intensity:DifferentSpread}(a-h), the evolutionary patterns of limit orders intensity when $s(0^-)=0.01$ is significantly different from the other cases with $s(0^-)\geq0.02$. As shown in Fig.~\ref{Fig:LOBResilency:Intensity:DifferentSpread}(j-l) and Fig.~\ref{Fig:LOBResilency:Intensity:DifferentSpread}(n-p), although the intensity curves of buy limit orders and sell limit orders do not fully overlap when $s(0^-)\geq0.02$, their differences are not marked. Both curves in each plot slowly decay to their normal levels within about 30 minutes. This means that the effective market orders produce \emph{symmetrical} stimulus to Type-5 buy limit orders and Type-11 sell limit orders under the case of $s(0^-)\geq0.02$. However, for the case of $s(0^-)=0.01$, the intensity curves of buy limit orders and sell limit orders show significant discrepancy. Fig.~\ref{Fig:LOBResilency:Intensity:DifferentSpread}(i) shows that, within 3 minutes after effective buy market orders, the intensity of Type-5 orders is higher than that of Type-11 orders. Fig.~\ref{Fig:LOBResilency:Intensity:DifferentSpread}(m) illustrates that there are more Type-11 sell limit orders than Type-5 buy limit orders after effective sell market orders. This suggests that effective market orders produce \emph{asymmetrical} stimulus to Type-5 orders and Type-11 orders under the case of $s(0^-)=0.01$. Specifically, effective buy market orders attract more buy limit orders at best bid (Type-5 orders), while effective sell market orders attract more sell limit orders at best ask (Type-11 orders). These observations can be used to explain the price continuation behavior after less-aggressive effective market orders (Type-3 and Type-9 orders). Indeed, the intensity curves in Fig.~\ref{Fig:LOBResilency:Intensity:DifferentSpread}(i,m) and in Fig.~\ref{Fig:LOBResilency:Intensity}(j,l) share similar patterns.

\section{Conclusions}

In this paper, we quantify and compare the resiliency of limit-order book after the submission of different types of effective market orders, using ultrahigh frequency data sets from the Chinese stock market. The orders are classified by their aggressiveness and the resiliency is analyzed based on three dimensions, namely the bid-ask spread, the LOB depth, and the intensity of incoming orders. We adopt a traditional approach to filter the intra-day seasonality of these indicators and then take average of the same type orders. Our results suggest that the evolutionary consistency between bid-ask spread/depth and order intensity.

First, the relative spread before the arrival of effective market orders is approximately minimal in almost all cases, which indicates that submitting market orders is more likely when the liquidity is high. The resiliency patterns of bid-ask spread show obvious diversity among different types of market orders. However, they all can return to the sample average within 20 best limit updates.

Second, effective market orders are prone to take place when the same side depth is high and the opposite side depth is low. This phenomenon is especially significant when the initial spread is 0.01 CNY (1 tick). After a market order shock, the LOB depth will also recover within about 20 best limit updates. Furthermore, the LOB resiliency behavior of depth is quite symmetric for Type-1 order shock vs. Type-7 order shock, Type-2 order shock vs. Type-8 order shock, and Type-3 order shock vs. Type-9 order shock.

Third, aggressive market orders do attract more placements of limit orders. After an aggressive buy (sell) market order shock, sell (buy) limit orders contribute more than buy (sell) limit orders to the resiliency of spread. In other words, the price resiliency behavior is dominant after Type-1 and Type-7 orders. However, after a relatively less-aggressive market buy (sell) order shock, limit buy (sell) orders contribute more to the resiliency of spread than limit sell (buy) orders. This means that, after Type-2, Type-8, Type-3 and Type-9 orders, the price continuation behavior is dominant. Moreover, effective market orders produce \emph{symmetrical} stimulus to limit orders when the initial spreads are larger. On the contrary, effective market orders produce \emph{asymmetrical} stimulus to limit orders when the initial spreads equal to 0.01 CNY. In this case, effective buy market orders excite more buy limit orders and effective sell market orders attract more sell limit orders. This observation is important because it can reasonably explain the price continuation behavior after less-aggressive effective market orders (Type-3 and Type-9 orders).

The analysis of LOB resiliency can be applied to improve the estimation of the transient or permanent price impact \cite{Bouchaud-Gefen-Potters-Wyart-2004-QF,Bouchaud-Kockelkoren-Potters-2006-QF,Farmer-Gerig-Lillo-Mike-2006-QF,Besson-Lasnier-2015}, to solve the optimal trade execution problem more precisely \cite{Alfonsi-Fruth-Schied-2010-QF,Obizhaeva-Wang-2013-JFinM}, and to design and calibrate computational models for order-driven markets \cite{Mike-Farmer-2008-JEDC,Gu-Zhou-2009-EPL,Li-Zhang-Zhang-Zhang-Xiong-2014-IS,Sornette-2014-RPP}.

\section*{Acknowledgments:}

This work was partially supported by the National Natural Science Foundation of China (71501072, 71532009, 71571121), the China Postdoctoral Science Foundation (2015M570342), and the Fundamental Research Funds for the Central Universities.

\section*{References}

\providecommand{\newblock}{}

\end{document}